\documentclass[draft]{agujournal2019}
\usepackage{subcaption}
\usepackage{amsmath}
\usepackage{comment}
\usepackage{lineno}
\usepackage{soul}
\usepackage{comment}
\usepackage{tabularx}
\usepackage{array}

\newcounter{SuppCount}
\newcommand{\SuppRef}[1]{%
	\stepcounter{SuppCount}%
	#1\theSuppCount
}
\newcommand{\eawaf}{East Atlantic }
\newcommand{\gal}{Param}
\newcommand{\comorph}{CoMorph}
\newcommand{\ral}{Explicit}
\newcommand{\dd}{\textrm{d}}
\newcommand{\hartL}{$-|V_T|^L$}
\newcommand{\hartU}{$-|V_T|^U$}

\draftfalse

\journalname{Journal of Advances in Modelling Earth Systems}

\begin{document}

\title{Dynamics of East Atlantic seed vortex populations in global km--scale models}
\authors{Ben Maybee\affil{1}, Francesca Morris\affil{2}, Juliane Schwendike\affil{1}, Ashar Aslam\affil{1}, Calum Scullion\affil{3}, Richard W.~Jones\affil{3}, Dasha Shchepanovska\affil{3}, and Kevin I.~Hodges\affil{4,5}}

\affiliation{1}{Institute for Climate and Atmospheric Science, University of Leeds, Leeds, UK}
\affiliation{2}{School of Geography and the Environment, University of Oxford, Oxford, UK}
\affiliation{3}{Met Office, Exeter, UK}
\affiliation{4}{National Centre for Atmospheric Science (NCAS), University of Reading, Reading, UK}
\affiliation{5}{Department of Meteorology, University of Reading, Reading, UK}
\correspondingauthor{Ben Maybee}{b.w.maybee@leeds.ac.uk}

\begin{keypoints}
\item A global km--scale, explicit convection model generates fewer, weaker Atlantic tropical cyclones than coarser, parameterised counterparts.
\item Simulations’ seed characteristics are equivalent: differences in TC activity arise from seed interactions with different mass flux profiles.
\item Mass flux profile characteristics are explained by biases in column moisture and Mesoscale Convective System structure.

\end{keypoints}

\begin{abstract} 
    Africa is the primary source of cyclonic vortices over the tropical Atlantic. Over both land and sea, these vortices are entwined with deep convective activity, with the majority being African Easterly Wave troughs. Their convective interactions have downstream impacts, since the same vortices provide the seed population for Atlantic basin tropical cyclone (TC) genesis. Understanding the dynamics of East Atlantic seed populations, particularly the processes that distinguish vortices which undergo cyclogenesis, is crucial for understanding the formation of Atlantic hurricanes and model representations of their populations. Here we investigate these questions in three one--year, atmosphere--land global km--scale Met Office Unified Model simulations. We use objective tracking algorithms to independently identify seed vortices, easterly waves, TCs, and Mesoscale Convective Systems (MCSs), benchmarking against reanalysis and satellite observations. 
    The simulations display equivalent continental and offshore seed vortex frequencies, however the highest--resolution, explicit convection simulation produces fewer, weaker hurricanes. Parameterised convection counterparts show stronger vortex amplification of seeds crossing the West African coastline, and more hurricanes form offshore. In the explicit simulation, we identify a failure to maintain strong vertical mass flux profiles experienced by seeds as the primary cause of weak vortex development, finding no low--level circulation development offshore about seed vortices. 
    Using MCS tracks, we show that systematic differences in convective organisation between the simulations can explain the differences in mass flux profiles, and thus vortex evolution. Deficiencies in the explicit simulation stem primarily from underestimation of MCS anvils and stratiform rainfall, a long--standing systematic bias shared with other explicit convection models.
\end{abstract}

\section*{Plain Language Summary}
    Tropical cyclones typically form from pre-existing atmospheric disturbances named seed vortices. These seeds are mesoscale regions of relatively weak rotation; tropical cyclones form from a seed when environmental factors or local interactions act to develop a strong, near--surface circulation with rapidly rotating winds. For North Atlantic hurricanes, most seed vortices are troughs of African Easterly Waves, mid-level vortices generated within the West African monsoon which propagate offshore from continental Africa. The evolution of East Atlantic seed vortices, and their propensity to develop into hurricanes, is strongly influenced by atmospheric moisture and interactions with large, but sub--vortex scale, organised deep convective systems. In this paper we study the processes which control seeds’ evolution in new, year--long global models run with a km--scale grid representing the atmosphere. In a simulation where deep convection is explicitly represented, we find that the offshore growth of seed vortex circulations is much weaker than in reanalysis and simulations which parameterise (approximate) convective motion. There are subsequently fewer, weaker Atlantic hurricanes in the explicit simulation. We show that offshore biases in column moisture, and the size, intensity and relative location of organised convective systems, are a key cause of the differences in seed vortex evolution between the simulations.
    

\section{Introduction}

Tropical cyclogenesis depends on large--scale environmental conditions and the availability of precursor seed vortices. Seeds are pre--existing compact regions of atmospheric vorticity, typically without the canonical full depth warm--core, axisymmetric structure of a tropical cyclone (TC; \citeA{Wang2025definition}). In the North Atlantic (NATL) and East North Pacific basins, the majority of available seeds originate over or near Africa: approximately 60\% of North Atlantic TCs, and 50\% of East Pacific TCs, are associated with African Easterly Waves (AEWs; \citeA{Russell2017revisiting,Du2025global}).

AEWs are extensively studied synoptic--scale waves that propagate in the mid/lower troposphere across central and western North Africa --- for a recent review, see \citeA{Nunez2026review}. The waves exist as barotropic--baroclinic disturbances of the African Easterly Jet (AEJ), associated with a reversal in the meridional potential vorticity (PV) gradient \cite{Pytharoulis1999low,Hsieh2007study}, and propagate in a northern and southern track relative to the AEJ at roughly 850 and 650\,hPa, respectively \cite{Thorncroft2001african}. The southern track AEWs in particular are intimately coupled to organised moist, deep convection in the form of Mesoscale Convective Systems (MCSs; \citeA{Mekonnen2006analysis,Semunegus2017characterization}). MCSs exhibit both convective and stratiform regions \cite{Janiga2014convection} and thus provide significant mid--level diabatic heating and PV sources, thereby driving AEW growth \cite{Russell2020potential} and playing an intrinsic role in AEW maintenance \cite{Berry2012african,Tomassini2017interaction,Russell2020african}. 

Offshore propagation of southern track AEWs provides a TC seed vortex population from which some ``developer vortices'' undergo cyclogenesis in the Atlantic basin's Main Development Region (MDR; \citeA{Thorncroft2001african,Hopsch2006west}). However, there is no correlation between annual hurricane counts and AEW numbers \cite{Russell2017revisiting}, the significant relationship instead being with lower--tropospheric eddy kinetic energy around the west coast of Africa. This points to the importance of vortex amplification and convective--interactions, rather than population size, in the frequency of East Atlantic tropical cyclogenesis. Indeed, sensitivity experiments suppressing AEW entry into the Atlantic basin show no significant decrease in NATL TC activity \cite{Patricola2018response,Danso2022influence,Bercos2023effects}.  Other known NATL seed vortex sources are frontal low--pressure systems and inter--tropical convergence zone vortex breaking \cite{Kouski2026influence}, and equatorial waves play important roles in favouring cyclogenesis \cite{Yang2018linking,Feng2023equatorial,Du2025global}. However, NATL TCs without AEW origins experience less favourable environmental conditions, make landfall further north, and develop weaker maximum winds than AEW--origin systems, resulting in lower economic impacts \cite{Bercos2024characteristics}.

The development of seed vortices into TCs can be caused by MCS interactions. MCS anvils show significant top--heavy vertical mass flux profiles due to their extensive stratiform rainfall regions \cite{Mapes1995diabatic,Raymond2014tropical} and radiative feedbacks \cite{Luschen2026stratiform}. The convective systems' mass flux profiles amplify mid--level vortices through vortex stretching and if maintained, further cause vortices to develop low--level circulations which interact with oceanic latent heat fluxes and undergo cyclogenesis \cite{Bister1997genesis,Raymond2007theory,Raymond2007evolution,Schwendike2010convection}. The likelihood of AEWs developing into TCs is observed to increase with coupling and collocation to large MCS anvils \cite{Leppert2013relation,Nunez2020wave}. This process chain is complementary to the frequently invoked ``marsupial pouch'' paradigm of genesis \cite{Dunkerton2009tropical,Wang2014characteristics,Asaadi2016dynamics}, but emphasises the interactions between convection and mid--level precursor vortices \cite{Raymond2014tropical}. For AEWs, offshore merger events between the mid and lower--level circulations of southern and northern track vortices, respectively, provide a further mechanism for vortex development which contributes to a significant minority of TCs forming from AEWs \cite{Chen2014relation,Duvel2021vortices,Jonville2025distinguishing}. 

Tropical cyclones, AEWs and MCSs are all phenomena characterised by convective scale--interactions, and thus are generally poorly represented by coarse--grid global climate models, which rely on parameterisations of convection \cite{Martin2015representation,Roberts2020impact,Slingo2022ambitious}. Even reanalyses struggle to accurately capture TC intensity, and thereby frequency, for the same reason \cite{Hodges2017well}.  The mutual interaction of these systems is therefore a prime opportunity for nascent large--domain km--scale, convection permitting (CP) simulations. Multiple year--long and even decadal global CP simulations now exist \cite{Segura2025nextgems} and are the subject of global co--ordinated assessments such as DYAMOND (DYnamics of the Atmospheric general circulation Modeled On Non-hydrostatic Domains; \citeA{Stevens2019dyamond,Takasuka2024protocol}). Explicit simulation of West African monsoon convection has remote influences across the entire Atlantic basin \cite{Pante2019resolving}, making global CP models some of the few suitable candidates for studying the full lifecycle of NATL TCs and their precursor vortices at kilometre scale.

Existing results indicate that global CP models improve the representation and intensity of TCs \cite{Judt2021tropical,Baker2024realism} and the lifecycles of MCSs \cite{Feng2025mesoscale}. While not convection permitting, analysis of TCs in large ensembles up to 25\,km resolution shows that decreasing model grid spacing improves TC intensity and frequency, particularly in the North Atlantic \cite{Roberts2015tropical,Roberts2020impact}. Over Africa, regional CP modelling significantly improves the representation of the West African monsoon \cite{Marsham2013role,Birch2014seamless}, including AEW propagation and intensity versus observations \cite{Tomassini2017interaction}, and MCS responses to environmental controls \cite{Maybee2024wind}. However, CP models are also afflicted by biases, including systematic biases towards convective rainfall in MCSs \cite{Feng2025mesoscale}; model--formulation dependent biases in TC structure \cite{Judt2021tropical}; weak TC activity globally versus reanalysis in specific models \cite{Emlaw2026understanding}; and very high relative costs and computational requirements \cite{Schar2020kilometer}.

In this article we use convective scale--interactions in the East Atlantic as a detailed case for investigating whether global CP models are able to realistically simulate interactions between mesoscale phenomena. North Atlantic hurricane activity is a key upscale model metric that stands to benefit from improvements in such interactions, making it crucial to assess the potential benefit offered by global CP modelling. By tracing seed vortex, MCS and TC evolution across large domains for a full season (Sec.~\ref{sec:methods}), we also aim to identify and study the physical processes that drive vortex development in greater comprehensive detail than is possible in reanalysis and observational case studies. To this end, in Sec.~\ref{sec:results} we explore differences in the West African monsoon, East Atlantic seed vortex populations and NATL TC activity in three global Unified Model simulations. Surprisingly, we find hurricane activity in a CP simulation is far weaker than in coarser--grid parameterised counterparts, despite all simulations having equivalent seed vortex frequencies. In Sec.~\ref{sec:processes} we identify key differences in seed vortex evolution and development mechanisms, isolating differences in vertical mass flux profiles stemming from coastal MCS characteristics (Sec.~\ref{sec:mcs}). We discuss the role of cascading biases in causing simulated TC activity in Sec.~\ref{sec:discuss}, before presenting our conclusions (Sec.~\ref{sec:conclusions}).


\section{Data and Methods}
\label{sec:methods}

We are interested in the dynamics of seed vortices over the East Atlantic ocean and continental West Africa during the peak North Atlantic hurricane season. We therefore define an \eawaf study region, taken as -45$^\circ$--15$^\circ$E by 0$^\circ$--25$^\circ$N (red box in Fig.~\ref{fig:tctracks}), and restrict to the period between June and October (inclusive). The oceanic part of our study area overlaps significantly with the MDR for Atlantic TCs.

\subsection{Data}

Our study focuses on three global, atmosphere--land only km--scale simulations from 20/01/20 to 01/03/21 performed with the non--hydrostatic, semi--Lagrangian Met Office Unified Model (MetUM, \citeA{Brown2012unified}) as part of the DYAMOND3 project \cite{Takasuka2024protocol}. Each simulation is driven by daily observed Operational Sea Surface Temperature and Ice Analysis (OSTIA) SSTs and sea--ice \cite{Donlon2012operational};  uses a stretched vertical grid with 85 levels up to 85 km (50 in the troposphere); outputs hourly surface and 3--hourly profile diagnostics; and is coupled to a land--surface explicitly represented by the JULES model \cite{Best2011joint}. All models use the same greenhouse gas forcings, with aerosol and dust climatologies generated with a full Earth system model.

The three simulations differ in their grid spacing and treatment of deep convection. \textit{\gal{}} and \textit{\comorph{}} both use a horizontal n1280 grid ($\sim$16$\times$10\,km spacing at the equator) and a convective parameterisation scheme. \gal{} uses the GAL9.0 Met Office science configuration that includes the same CAPE--closure mass flux parameterisation and turbulent mixing schemes found in the current global MetUM operational forecasting suite \cite{Willett2026met}. This configuration also includes the single moment Wilson--Ballard microphysics and PC2 prognostic cloud schemes \cite{Wilson1999microphysically,Wilson2008pc2}. \comorph{}, in contrast, deploys the more experimental CoMA9\_TBv1p2 configuration \cite{Lavender2026comorph}. This configuration is similar to GAL9.0 but makes use of the new CoMorph mass flux convection scheme built and tested by the Met Office \cite{Lock2024performance}, with slight modifications to GAL9.0 physics schemes to enable simulations in the convective ''grey--zone'' \cite{Lavender2026comorph}. Although originally developed for use at $\sim$5 km resolution, inclusion of a grid--scale dependent entrainment formulation gives an aspect of scale--awareness for the $\sim$10\,km \comorph{} simulation used here.

\textit{\ral{}} is a CP simulation run at global n2560 horizontal resolution ($\sim$8$\times$5\,km spacing at the equator) using the RAL3.3 science configuration developed for regional forecasting and climate applications \cite{Bush2025third}. The standard configuration includes CASIM double moment microphysics \cite{Field2023implementation}; a bimodal large--scale cloud parameterisation scheme \cite{vanWeverberg2021bimodal}; a blended boundary layer scheme \cite{Boutle2016london}; and scale--weighted 1D \cite{Lock2000new} and 3D Smarogrinsky turbulent mixing schemes \cite{Bush2025third}. Following identification of top of atmosphere radiation biases in previous CP global simulations, the RAL3.3 configuration used for global simulations here has retuned cloud radiative properties and a reduced cloud droplet number concentration to bring the global energy balance inline with CERES radiation budget observations \cite{Wielicki1996clouds}.

A key aspect of the models is that they are year--long, free--running simulations forced only by observed SSTs, and thus are not intended to provide faithful reflections of observations. In 2020 the SSTs were anomalous and drove an over--active NATL TC season relative to climatology \cite{Klotzbach2022hyperactive}, further complicating verification of the simulations. Nevertheless, observational benchmarking is essential. As all simulations are km--scale, we expect all to provide climatologically realistic representations of the seed vortex populations, with relatively strong TC activity given the favourable SSTs. Based on prior literature alone, we hypothesise that \ral{} will show the most realistic seed vortex and TC activity. To measure this, we use reanalysis and satellite observations to provide both seasonal (2020) and climatological benchmarks where possible. We primarily rely on ERA5 reanalysis spanning the 1979---2024 satellite data--assimilation era \cite{Hersbach2020era5}. We also use the 2001---2024 GPM IMERG V07 precipitation (only available from 2001; \citeA{Huffman2023integrated}), NOAA CPC/NCEP global merged infrared brightness temperature ($T_b$) data \cite{Janowiak2017ncep}, and the International Best Track Archive for Climate Sterwardship (IBTrACS) catalogue of TC best--track data complied by tropical warning centres \cite{Knapp2010international}.

\subsection{Seed vortex and TC tracking}

The TRACK algorithm \cite{Hodges1994general,Hodges1995feature} is used to objectively identify vortices. For seed vortex and TC tracking we use the implementation of the algorithm described in detail by \citeA{Hodges2017well}. Seed vortices are first identified globally between 60$^\circ$S and 60$^\circ$N from T63 spectrally filtered 6--hourly relative vorticity vertically averaged between 850 and 600\,hPa, labelled $\langle\zeta\rangle$, and excluding the first 0--5 total wavenumbers. Northern hemisphere tracks are generated by first identifying off--grid maxima in $\langle\zeta\rangle$ that exceed 5$\times$10$^{-6}$ s$^{-1}$, which are then linked into temporal tracks by using a nearest--neighbour first join and minimisation of a cost function \cite{Hodges1995feature}. The seed vortex population comprises all tracks which last for at least two days.

TCs are then identified as a subset of the seed vortex population which meet additional criteria applied to additional diagnostic fields that are added to the base tracks \cite{Hodges2017well}. The relevant added fields are the maximum T63 vorticity within a 5$^\circ$ search radius at levels 850 and 700---200\,hPa; minimum mean sea level pressure within a 5$^\circ$ radius; and maximum 10\,m winds within a 6$^\circ$ radius. TC tracks are required to begin within 30$^\circ$S---30$^\circ$N, and for a minimum of four consecutive time steps over ocean, both the T63 850\,hPa relative vorticity $\zeta$, and the difference in T63 850 and 200\,hPa $\zeta$, must be larger than 6$\times$10$^{-5}$ s$^{-1}$. This ensures the system is appropriately intense and has a warm core structure. A T63 vorticity centre must also exist at each level between 850 and 200\,hPa for the same four timesteps to ensure a coherent vertical structure \cite{Hodges2017well}. 

Beyond the robust identification process, TRACK is advantageous for our analysis for several reasons: firstly the T63 spectral filtering ensures track characteristics for MetUM simulations and ERA5 are coherent and sampled at a common horizontal resolution. Secondly, TRACK's TC tracks extend far beyond the period during which the system is categorised as a tropical depression or stronger, enabling communication between seed vortex and mature TC characteristics. We label those seeds which later go on to meet the additional TC criteria as developer vortices (DVs). Non--developer vortices (NDVs) never evolve into TCs. Since our focus is on seeds, ``genesis'' events always refer to the start of a seed track, not the point of intensification of a developer into a TC, i.e. cyclogenesis.

For all tracked vortices, we also calculate the Hart phase space parameters at each timestep:
\begin{equation}
-|V_T|^L = \frac{\partial\Phi(p)'}{\partial\log p}\bigg|_{p=900\textrm{hPa}}^{p=600\textrm{hPa}}\,, \qquad -|V_T|^U = \frac{\partial\Phi(p)'}{\partial\log p}\bigg|_{p=600\textrm{hPa}}^{p=300\textrm{hPa}}\,,
\end{equation}
where $\Phi(p)'$ is the difference between the maximum and minimum geopotential values  within a 5$^\circ$ search radius of the vortex centre at pressure $p$ \cite{Hart2003cyclone}. The Hart parameters encode the thermal wind response of a vortex, with positive values indicating a warm core in which $\Phi'$ decreases with height.

To calculate vertical profiles of vortex thermodynamic and dynamical fields, we must add a measure of vortex geometry to the point--tracking provided by TRACK. We ground our approach in the calculation of vortex circulation $\Gamma$, since circulation is defined as 
\begin{equation}
    \Gamma=\int\eta\,\dd S = \oint_C\mathbf{v}\cdot\dd\mathbf{\ell}\,,
    \label{eqn:circulation}
\end{equation}
where $\mathbf{v}$ is the horizontal velocity vector, $\eta=f+\zeta$ is the absolute vertical vorticity, and $C$ is a contour with unit element $\dd\mathbf{\ell}$ that bounds a vertically--oriented surface $S$. To calculate $\Gamma$, one must select an appropriate $C$, which can then be used as the domain for calculating other relevant fields.

Typically, studies of vortex circulation focus on case studies or regional domains and can thus sensibly subjectively choose a fixed contour over which $\Gamma$ is calculated \cite{Schwendike2010convection,Raymond2011vorticity,Morris2024synoptic,Morris2025closing}. In this study we must account for large seed vortex populations, and the potential for vortex interactions. We therefore identify contours objectively by applying the TRACK vorticity maxima--detection threshold (5$\times$10$^{-6}$\,s$^{-1}$) to mask the underlying T63 vertically--averaged vorticity fields. A unique vorticity footprint is then assigned to each vortex centre; where footprints merge between vortex centres, a geodesic nearest--neighbour criteria is used to divide them uniquely. The contour $C$ for a vortex is then taken as the Cartesian bounding box around the vortex $\langle\zeta\rangle$ footprint to ease integration. This procedure is repeated at all timesteps for all \eawaf vortices. 

For vortex--relative thermodynamic profiles we then take area means over vortex contours. In all simulations, vertical profiles for the grid--scale vertical mass flux $\rho w$, where $\rho$ is density, are obtained from explicit vertical velocities $w$ (only), calculated as the area integral over a vortex contour, and normalised by contour area. Note that each vortex contour $C$ can, and will, change for that vortex between timesteps, since the size of the region above the T63 threshold will fluctuate.

\subsection{AEJ and MCS identification}
\label{sec:mcs_criteria}

We also track MCSs in each dataset using the simpleTrack algorithm \cite{Stein2014three,Maybee2025how}. Storms are identified within the \eawaf region during JASO from the cloud brightness temperature ($T_b$) field as snapshots with a minimum area of 1000\,km$^2$ where $T_b<$-32$^\circ$C. Rainfall volumes and extremes are calculated for these tracks. We then filter MCS tracks using similar criteria to the MCSMIP intercomparison \cite{Feng2025mesoscale}. We restrict to only snapshots with areas larger than 15,000\,km$^2$. MCSs are then tracks which persist (at this threshold) for longer than 4 hours, and achieve a maximum lifetime area of 40,000\,km$^2$, a lifetime maximum rainfall rate above 10\,mm hr$^{-1}$, and a lifetime maximum rainfall volume above 20,000\,km$^2$ mm hr$^{-1}$. In the MetUM simulations, $T_b$ is derived from top--of--atmosphere outgoing longwave radiation. For observed MCS tracks, which we label OBS--MCS, we use the previously specified satellite products (CPC/NCEP and IMERGv7) from 2015 to 2024.

A fundamental component of the West African monsoon circulation is the AEJ, which should be identified from zonal winds on multiple pressure levels to account for variability in jet height \cite{Bercos2020relationship}. We use the vertical layer between 750 and 600\,hPa to include three MetUM diagnostic pressure levels. To assess the variability of the AEJ we adopt the method outlined by \citeA{Klotzbach2025remarkable}, identifying the minimum (i.e. strongest) 5--30$^\circ$N zonal wind over this vertical layer in a zonal average taken over -20$^\circ$--15$^\circ$E. We then take daily average latitudes and wind magnitudes to construct a time--series of jet characteristics for each dataset.


\section{Results: basic states}
\label{sec:results}

Our results are structured by scale, gradually zooming in to relevant local processes. In this section we begin with East Atlantic seasonal mean state, and identify the basic seed vortex and TC populations within the three simulated seasons. 

\subsection{\eawaf mean states}
\label{sec:mean_state}

\begin{figure}[h!]
    \centering
    \includegraphics[width=0.49\textwidth]{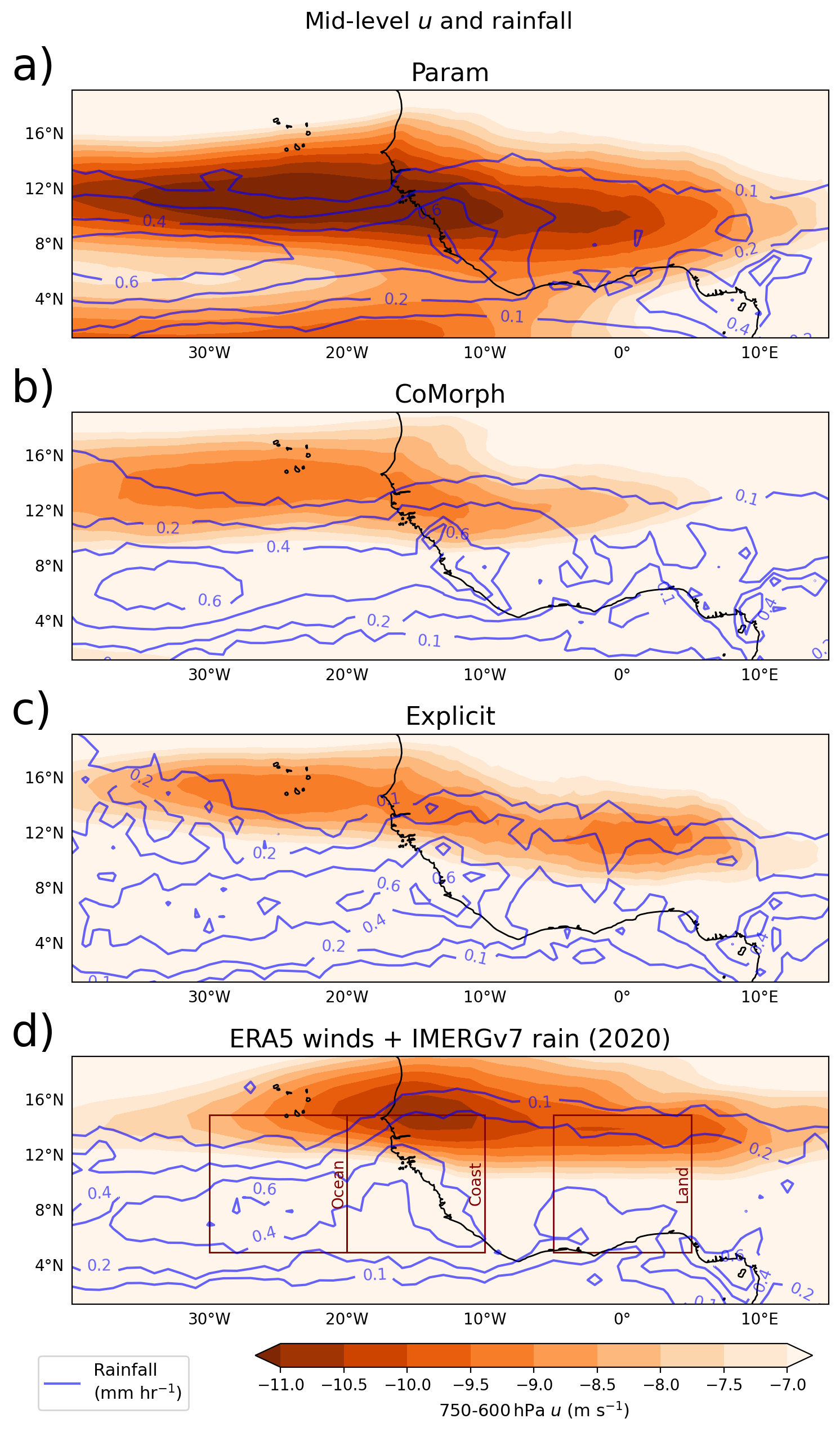}
    \includegraphics[width=0.49\textwidth]{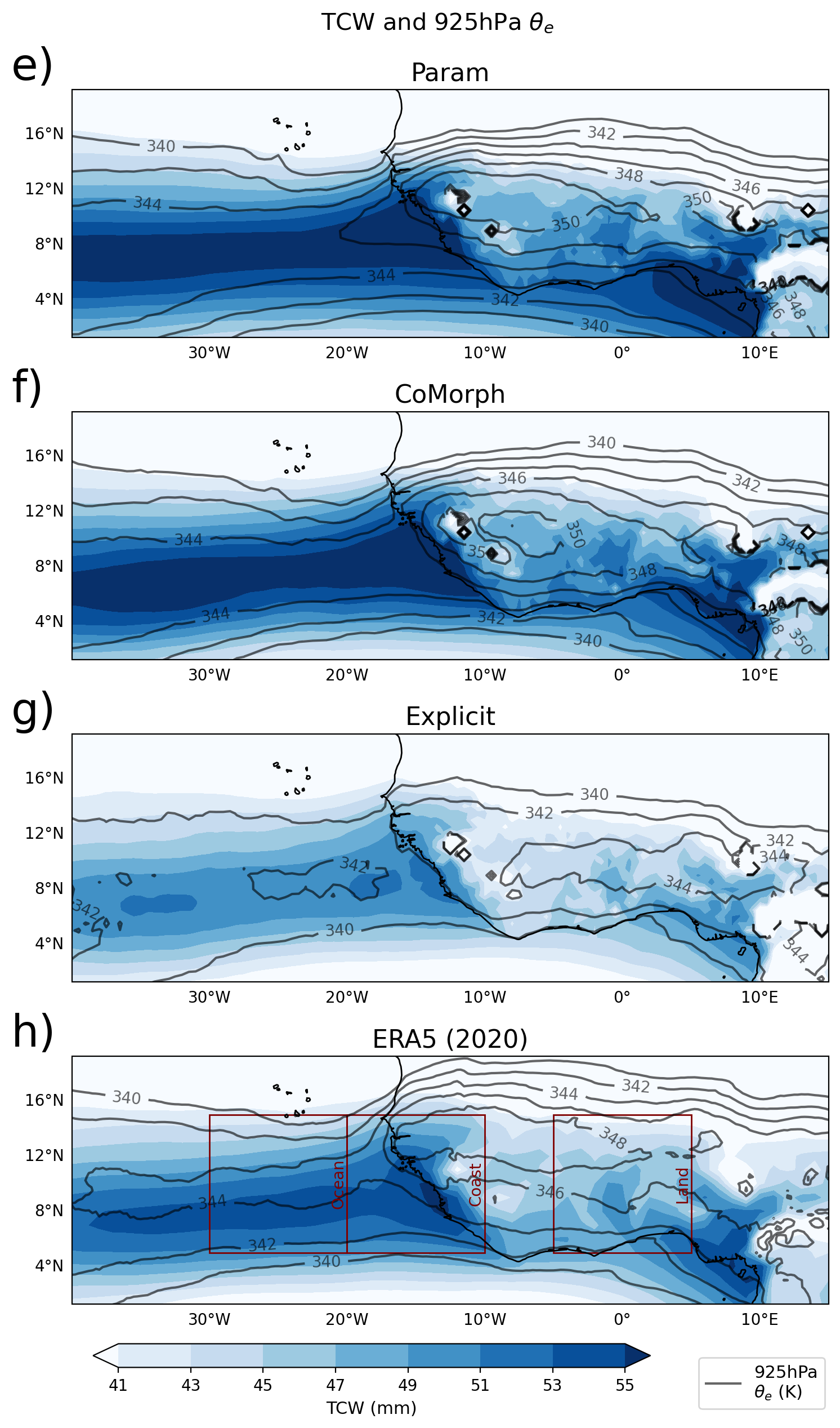}
    \caption{East Atlantic JJASO 2020 seasonal mean state. Panels \textit{(a---d)} show 750--600\,hPa zonal winds $u$ (shading) and rainfall (contours) for each MetUM simulation, and combined ERA5 winds and IMERGv7 rainfall, respectively. Repeated in panels \textit{(e---h)} for total column water (TCW, shading) and 925\,hPa equivalent potential temperature ($\theta_e$, contours). Going from west to east, red boxes on ERA5 panels show the (equally sized) Ocean, Coastal and Land subdomains used in this study.}
    \label{fig:wam_mean}
\end{figure}

Seed vortices in the East Atlantic basin are effectively mesoscale eddies within the West African monsoon circulation, with the AEJ providing their mid--level steering flow. Intra and inter--seasonal variability of the jet's position and strength are directly tied to variability in monsoon rainfall \cite{Grist2001study,Bercos2020relationship}, ensuring monsoon state plays a crucial role in modulating variability of tropical cyclogenesis. Figure~\ref{fig:wam_mean} shows fundamentally different representations of the monsoon across the simulations. The AEJ is notably stronger in \gal{} and extends far further west off the African continent (Fig.~\ref{fig:wam_mean}a), while there is a dry rainfall bias (contours) over land east of 5$^\circ$W versus IMERG observations (contours, \ref{fig:wam_mean}d). The jet in \ral{} (\ref{fig:wam_mean}c) has the closest position to the seasonal ERA5 mean (shading, \ref{fig:wam_mean}d), sitting relatively far north, but the observed jet's strength lies between that of \gal{} and \ral{}. The AEJ in \comorph{} (\ref{fig:wam_mean}b) extends too far west and is weaker again. Spatial rainfall patterns in \ral{} and \comorph{} compare well to IMERG, showing significant improvement versus \gal{} over land in particular, however the monsoon rains (contours) do not extend far enough north in all simulations. Climatologically, the mean AEJ state is weaker and further south than seen in 2020, and the monsoon does not extend so far north (Fig~S\SuppRef{}a). \ral{} shows a realistic representation of the climatological monsoon, but therefore underestimates the strength of the more unusual 2020 season.

\begin{figure}[t]
    \centering
    \includegraphics[width=\textwidth]{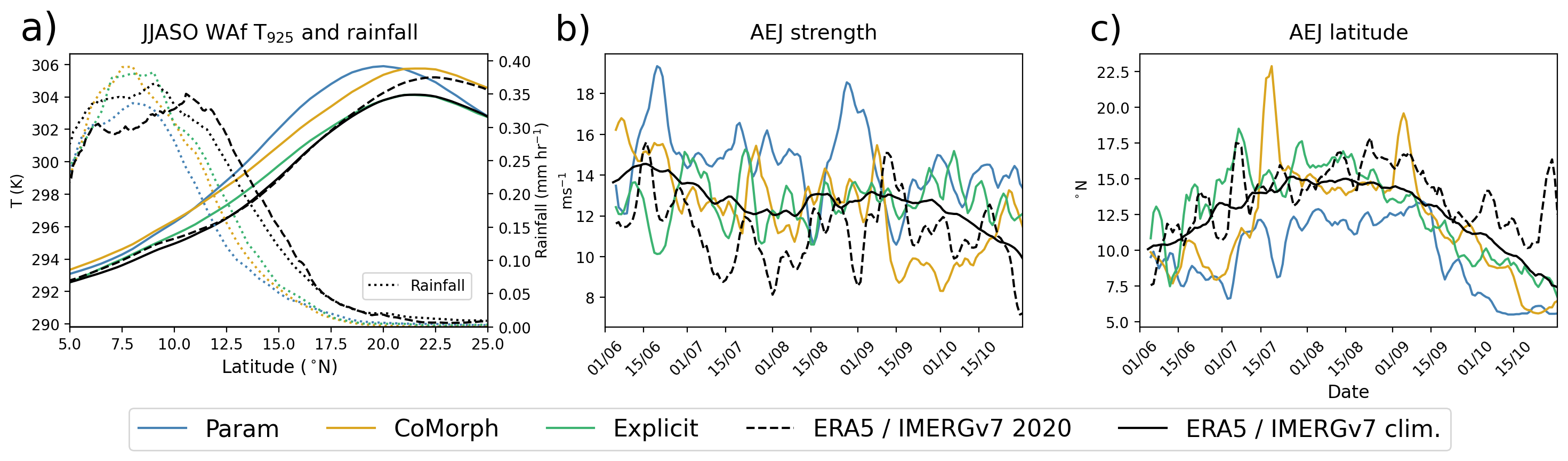}
    \caption{\textit{(a)} JJASO zonal mean 925\,hPa temperature (solid) and rainfall (dotted), with longitudinal average taken over -15$^\circ$-- 15$^\circ$E. \textit{(b)} Time series of daily mean AEJ strength, smoothed to a five day rolling mean, repeated in \textit{(c)} for AEJ latitude. In all panels, black dashed lines show 2020 ERA5 or IMERGv7 values, and black solid lines their climatologies. }
    \label{fig:aej_bits}
\end{figure}

The relative behaviour of the mean monsoon rains cannot be explained by moisture alone: both \gal{} and \comorph{} show a significant positive total column water (TCW) bias versus ERA5 (Figs.~\ref{fig:wam_mean}e,f,h), including over regions with deficient rainfall. Meanwhile \ral{} exhibits a dry TCW bias (Fig.~\ref{fig:wam_mean}g). For all simulations, TCW biases are most pronounced over the ocean. The significantly moister atmospheric columns in the parameterised simulations correlate with a moister boundary layer, here shown through 925\,hPa equivalent potential temperature ($\theta_e$) to also measure the simulations' warmer near--surface air temperatures (Fig.~\ref{fig:aej_bits}a). In reanalysis, the 2020 TCW and $\theta_e$ means are very similar to climatology, with the main difference being a narrow zonal band of moister, warmer air that stretches far offshore and is thus likely driven by SST patterns (Fig.~S1b). The climatologically more realistic inland rainfall and AEJ in \ral{}, despite a more favourable high--TCW, high--$\theta_e$ convective environment in \gal{}, is consistent with previously documented failings of parameterised models over West Africa in representing the monsoon circulation and AEJ feedbacks \cite{Marsham2013role,Birch2014seamless}. It is encouraging that the newer convective parameterisation scheme used in \comorph{} here results in a more realistic monsoon state. Offshore, the reduced influence of the AEJ, and the much moister atmospheres of the parameterised models, ensures their rainfall totals are more equivalent to \ral{} and observations.

The AEJ is fundamentally a response to the large--scale meridional temperature gradient between the dry, hot Sahara and moist Guinea coast \cite{Cook1999generation}. Figure~\ref{fig:aej_bits}a shows that the zonal mean temperature of the Sahara (i.e. $\sim$20$^\circ$N) is higher in \gal{} (blue line), and peaks further south, than the other models. This causes a stronger meridional temperature gradient, shifted further south, and thus a stronger AEJ which is located further south through thermal wind balance. The temperature gradients in \comorph{} (orange) and \ral{} (green), in contrast, are weaker, while their monsoon rainfall (dotted lines) extends further north. Figures~\ref{fig:aej_bits}b and~c show that in \gal{}, AEJ strength peaks in late June, when it also transitions north, reflecting a late monsoon onset date. \comorph{} and \ral{} are closer to the ERA5 seasonal (dashed black lines) and climatological cycles (solid black, Fig.~\ref{fig:aej_bits}c). The weaker seasonal mean winds in \comorph{} (Fig.~\ref{fig:wam_mean}b) reflect an early decline and southern shift of the jet (Fig.~\ref{fig:aej_bits}b), while in \ral{} the jet is frequently the farthest north during the first half of the season (Fig.~\ref{fig:aej_bits}c). The 2020 ERA5 AEJ behaves similarly but remained much further north than climatology from mid--August onwards. Despite appearing stronger than average in the spatial mean (Fig.~S1a), our daily metric shows the jet's strength was often weaker, suggesting the signal in Fig.~\ref{fig:wam_mean}d is dominated by its strong northwards excursion in early September (Fig.~\ref{fig:aej_bits}b,c).

\citeA{Marsham2013role} have previously illustrated that convective organisation drives an upscale modulation of monsoon circulation using regional convective--scale simulations. The meridional temperature gradient over West Africa is ultimately determined by the strength of the Saharan heat low. Overnight, northward propagating MCS cold pools ventilate the heat low and thus reduce the regional meridional temperature and pressure gradient \cite{Marsham2013role,Maybee2025how}. Systematic errors in the diurnal cycle of convection in parameterised models prevent this ventilation, leading to precisely the regional--scale errors seen here in \gal{} \cite{Garcia2013impact,Birch2014seamless}, suggesting that the representation of convection plays an important upscale role.

Briefly considering the wider tropical Atlantic basin, recall all simulations are driven by observed SSTs and thus share a favourable ocean state for tropical cyclogenesis, since anomalously warm summer Atlantic SSTs played a pivotal role in making 2020 the most active hurricane season on record \cite{Klotzbach2022hyperactive}. Further key favourable environmental conditions for cyclogenesis are weak vertical wind shear throughout the troposphere, high mid--level humidity, low saturation deficits, and high potential intensity \cite{Emanuel2022tropical}. Such environments are evident in the seasonal mean over the tropical Atlantic in all MetUM simulations (Fig.~S\SuppRef{}): the parameterised models have moister mid--levels and marginally lower saturation deficits, but all models show weak shear through the MDR, and potential intensities are much higher in \ral{}. 

\subsection{Seed vortex populations}
\label{sec:seeds}

\begin{figure}[t]
    \centering
    \includegraphics[width=\textwidth]{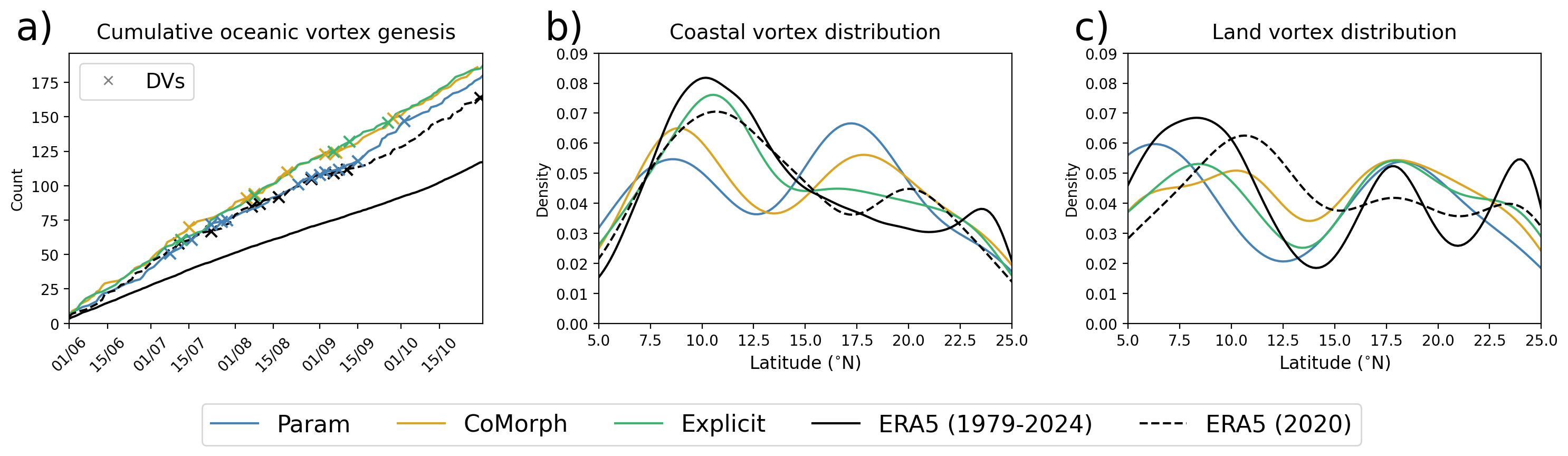}
    \includegraphics[width=\textwidth]{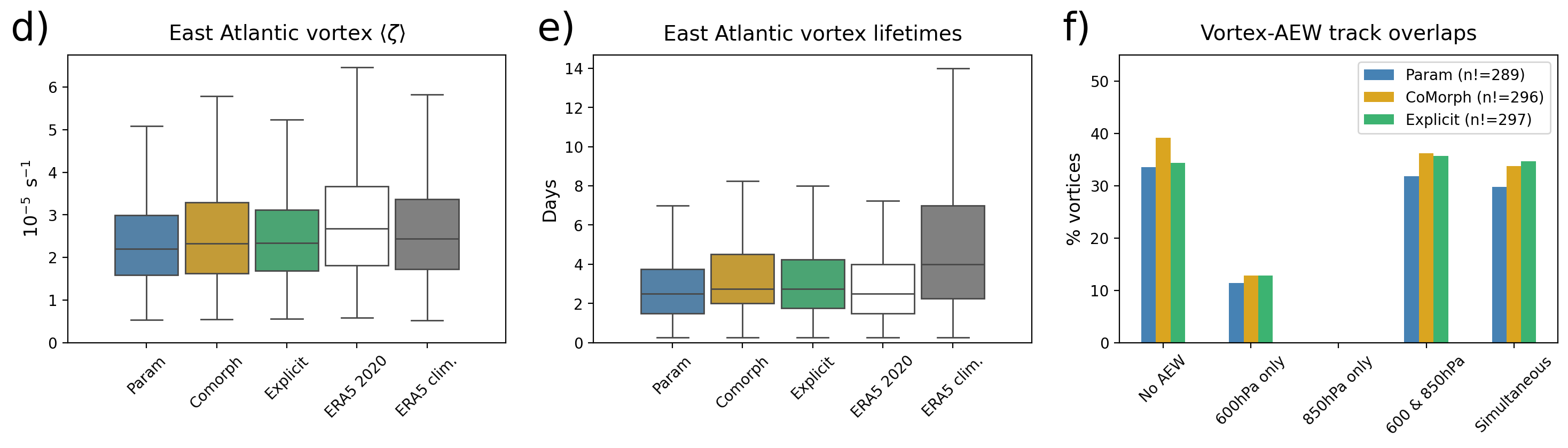}
    \caption{Seed vortex characteristics during JJASO. \textit{(a)} Time series of the cumulative count of \eawaf seed vortex genesis events over the ocean. Genesis here includes the coastal transition of a land--based vortex; genesis events for DVs are indicated by crosses. \textit{(b)} Kernel density estimate of daily mean vortex latitude in the Coastal longitude band (10$^\circ$-- 20$^\circ$W), fitted to data using a Gaussian kernel. \textit{(c)} Repeated for Land longitudes (-5$^\circ$-- 5$^\circ$E). Panels \textit{(d)} and \textit{(e)} then show distributions of domain lifetimes and lower--tropospheric vertically--averaged relative vorticity $\langle\zeta\rangle$, respectively, for all East Atlantic seed vortices. ERA5 2020 values have no shading, ERA5 1979--2024 climatology shaded grey. \textit{(f)} Bar charts of percentage of seed vortex tracks overlapping with a tracked AEW trough. At each timestep, an AEW trough at specified level and a seed vortex are considered to overlap if the AEW centre location lies within the seed vortex contour. Category ``600 \& 850\,hPa''refers to both AEW--level overlaps occurring independently at some instance during track lifetime, while ``Simultaneous'' refers to both occurring simultaneously. The n! values in legend specify numbers of unique seed vortices.}
    \label{fig:aej_seeds_overview}
\end{figure}

Turning to the basic state of the East Atlantic seed vortex populations, Fig.~\ref{fig:aej_seeds_overview}a shows the cumulative number of individual oceanic vortices identified by TRACK through the Atlantic hurricane season in each simulation. All MetUM simulations (coloured lines) have very similar oceanic seed activity, whether from offshore transport of land seeds or ocean--only genesis. Seed counts are highest in \ral{}, but the marginally lower counts in \gal{} are most similar to ERA5 in 2020 (dashed black line), especially before September. The ERA5 climatological mean seed frequency (solid black) is notably weaker than in 2020. Further including land vortices, in total we identify 289 seeds in \gal{}, 296 in \comorph{}, and 297 in both \ral{} and ERA5. Despite the equivalent magnitudes of the 2020 seed populations, there are only 5 DVs (crosses) in \ral{}, versus 12 in GAL9, 7 in \comorph{} and 9 in ERA5 during 2020. We focus on this key finding in the next subsection. There is some relation between DV genesis seasonality (crosses) and the AEJ, with late--August clusters of events in \gal{} and ERA5 coinciding with second peaks in jet strength (Fig.~\ref{fig:aej_bits}b). 

The zonal distribution of Coastal vortices in \ral{} (Fig.~\ref{fig:aej_seeds_overview}b) peaks $\sim$4$^\circ$ further north than \gal{}, closely following the ERA5 baseline and matching the more northerly AEJ in \ral{} compared to \gal{}. There is greater consistency in the zonal distributions south of the AEJ over land (Fig.~\ref{fig:aej_seeds_overview}c), however all MetUM models show stronger vortex activity than ERA5 north of 15$^\circ$N, both over land and at the coast. In the simulations and climatology, the majority of vortex genesis events occur east of our \eawaf domain, consistent with the genesis of AEW trains around orography east of the Sahel. Genesis events also cluster around the Cameroon highlands and the Guinea coast (Fig.~S\SuppRef{}a).

The distributions of seed vortex vorticities (Fig.~\ref{fig:aej_seeds_overview}d) and  lifetimes (Fig.~\ref{fig:aej_seeds_overview}e) are comparable between the simulations (colour shading), with similar median values and no systematic increase in \gal{} seed vorticity $\langle\zeta\rangle$. The ERA5 2020 seed vortex population (no shading) is skewed to stronger vorticities and marginally shorter lifetimes than the MetUM simulations. All the 2020 populations show seed lifetimes which are significantly shorter than in climatology (grey shading), which has a median of 4.00 days versus 2.75 in \ral{} and \comorph{} and 2.50 in ERA5 during 2020. This difference is primarily because of longer--lived land vortices, since most ocean vortices leave the study region's western boundary and thus dissipate artificially (Fig.~S3b). In ERA5 climatology we also find a larger proportion of land--ocean vortices, at 50\% of the population, versus around 25\% in the 2020 simulations and reanalysis.

The majority of seed vortices identified in the simulations are AEW troughs. This is strongly implied by the location of the latitudinal peaks of the vortex distributions (Figs.~\ref{fig:aej_seeds_overview}b,c), which match the climatological location of observed AEW trains \cite{Nunez2026review}. To confirm the identity of seeds we independently ran TRACK on 600\,hPa and 850\,hPa T63 vorticity using the method detailed in \citeA{Thorncroft2001african}, corresponding to southern and northern track wave levels, respectively. Spatial distributions of identified AEW troughs are shown in Fig.~S\SuppRef{}. Figure~\ref{fig:aej_seeds_overview}f then quantifies the percentage of seed vortices for which a tracked AEW centroid lies within the vortex contour (overlaps). In all simulations over 60\% of seed vortex tracks overlap with a tracked 850\,hPa or 600\,hPa trough and may thus be considered AEWs. These seeds are all southern track AEWs: there are no systems which only overlap with a 850\,hPa wave trough (Fig.~\ref{fig:aej_seeds_overview}f). Overlaps with 850\,hPa troughs instead mostly occur at the same time as a mid--level vortex (``Simultaneous'' category). This simple labelling cannot distinguish between vortex deepening and AEW track merger events as the physical mechanism underpinning low--level circulation development, but does demonstrate that this occurs in around 30\% of seed vortex tracks in all simulations.

\subsection{TC populations}
\label{sec:tcs}

\begin{figure}[t]
    \centering
    \includegraphics[width=\textwidth]{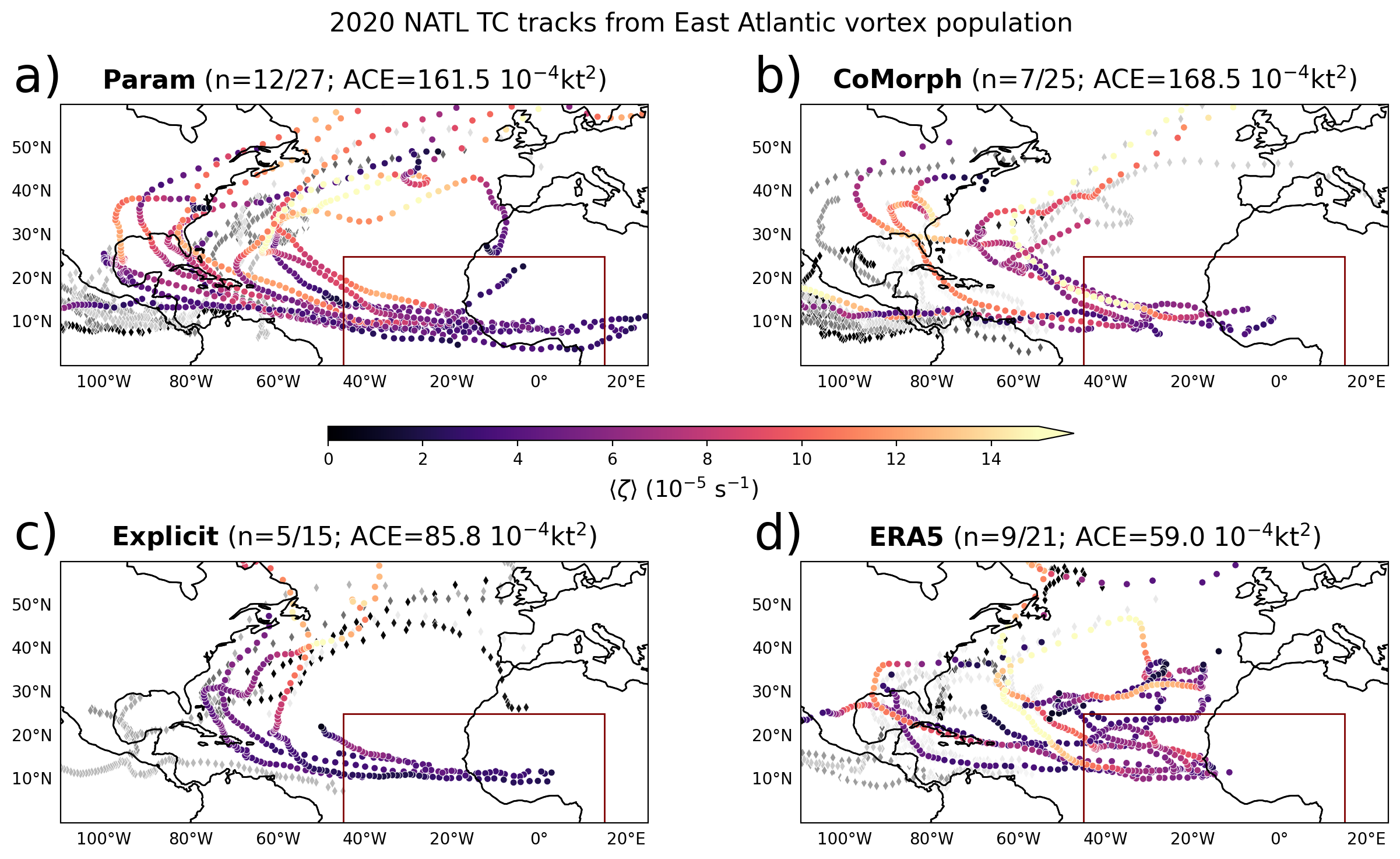}
    \caption{Full tracks for all 2020 NATL basin vortices which develop TC status in \textit{(a)} \gal, \textit{(b)} \comorph, \textit{(c)} \ral{} and \textit{(d)} ERA5. Tracks with genesis in the \eawaf study region (red box) are shaded by vertically averaged lower--tropospheric vorticity $\langle\zeta\rangle$; tracks with genesis elsewhere in the basin shaded grey. Numbers in parentheses record the fraction of system counts, $n$, for NATL basin TCs which develop from the \eawaf region, and total basin seasonal ACE.}
    \label{fig:tctracks}
\end{figure}

Figure~\ref{fig:aej_seeds_overview}a already shows that \ral{} has fewer than half of the number of East Atlantic DVs which later undergo tropical cyclogenesis than \gal{}, despite having no relative shortage of seeds. Taking a full--basin view, Fig.~\ref{fig:tctracks} plots the 2020 full--season TC tracks for each simulation and ERA5. Tracks of TCs which developed from a seed generated in the \eawaf region, shown by red boxes in panels, are shaded by their vorticity $\langle\zeta\rangle$. The tracks immediately reveal a dearth of Atlantic TCs in \ral{} (Fig.~\ref{fig:tctracks}c), with 15 systems in total versus 27 in \gal{} (\ref{fig:tctracks}a) and 25 in \comorph{} (\ref{fig:tctracks}b). For reference, in IBTrACS observations there were 29 TCs (Fig.~S\SuppRef{}a). Applying TRACK to reanalysis (ERA5 in Fig.~\ref{fig:tctracks}d, and JRA--3Q in Fig.~S5b) further supports the conclusion that \ral{} suffers an underestimated NATL TC population: reanalyses underestimate TC intensities, and thus system counts versus IBTrACS due to thresholding \cite{Hodges2017well}. We have also confirmed that this result is insensitive to the choice of tracking algorithm, finding consistent results with the TempestExtremes framework (\citeA{Ullrich2021tempestextremes}; not shown).

In all MetUM simulations, systems follow physically realistic tracks; the canonical AEW pathway leading to recurvature west of the MDR is clearly visible in Fig.~\ref{fig:tctracks}. In all datasets except \ral{} at least one track has $\langle\zeta\rangle>10^{-4}$\,s$^{-1}$ before leaving the \eawaf region. In \ral{} cyclogenesis occurs further west and north than in other datasets, and system vorticities remain significantly lower throughout their lives. The small, weak TC population in \ral{} gives a total accumulated cyclonic energy (ACE) of 85.8$\times$10$^{4}$\,kt$^2$, nearly half of that in \gal{} and \comorph{} (161.5$\times$10$^{4}$\,kt$^2$ and 168.5$\times$10$^{4}$\,kt$^2$, respectively). ACE is a measure of seasonal TC activity that includes both system counts and intensities, thus providing a valuable single quantification of the inter--simulation differences apparent in Fig.~\ref{fig:tctracks}. Note we follow the standard convention of only calculating ACE for TCs with a maximum wind speed in excess of 34\,kt. On this measure the IBTrACS value of 187.1$\times$10$^{4}$\,kt$^2$ should be taken as the observational benchmark rather than ERA5, but we emphasise that seasonal NATL ACE exhibits significant interannual variability (e.g. Fig.~28 in \citeA{Bell2000climate}).  

When benchmarking the simulations, it is crucial to bear in mind they are free--running (from February 2020) and thus expected to diverge from reality: the purpose of the comparison is rather to highlight the variability in inter--model regional spread. The surprise here is that the simulation with the most physically realistic West African monsoon, \ral{}, shows the weakest TC population, in contrast to previous work showing that improved monsoon state at finer grid spacing in parameterised models leads to improved NATL TC variability \cite{Roberts2015tropical,Roberts2020impact}. Seasonal TC counts for primary basins globally suggest that the weak NATL TC activity in \ral{} is a regional bias, since the simulation generates numerous, intense TCs in the West North Pacific and Indian Ocean (Figs.~S5c,d). The TC tracks and counts further point to properties of the \eawaf seed vortex population as the origin of the weak \ral{} Atlantic TC population: the zonal Atlantic AEW train and MDR activity are both far less active than other datasets (Fig.~\ref{fig:tctracks}), and the TC count in the East North Pacific basin is also notably underestimated (Fig.~S5c). AEWs continue to play a key role in cyclogenesis in this basin \cite{Chen2008north,Serra2010tracking}, and indeed we find no Pacific TC seeds emanating from West Africa in \ral{}.

\section{Results: process analysis}
\label{sec:processes}

The three km--scale MetUM simulations show very different NATL TC activity despite significant differences in \eawaf seed vortex frequency and mean state over the full ocean basin. There are significant mean state differences in the simulations' West African monsoon, but the highest resolution, convection permitting simulation, \ral, that shows the most physically realistic monsoon also produces far fewer TCs than parameterised counterparts. If the number of seeds is the same (Fig.~\ref{fig:aej_seeds_overview}), and the large--scale mean state of the Atlantic is similar (Fig.~S2), then the simulations must be evolving their seed populations differently.  We therefore now turn to the mesoscale dynamics that govern the intensification and selection of seed vortices that develop into TCs.

\subsection{Seed vortex evolution}
\label{sec:vortices}

\begin{figure}[t]
    \centering
    \includegraphics[width=0.49\textwidth]{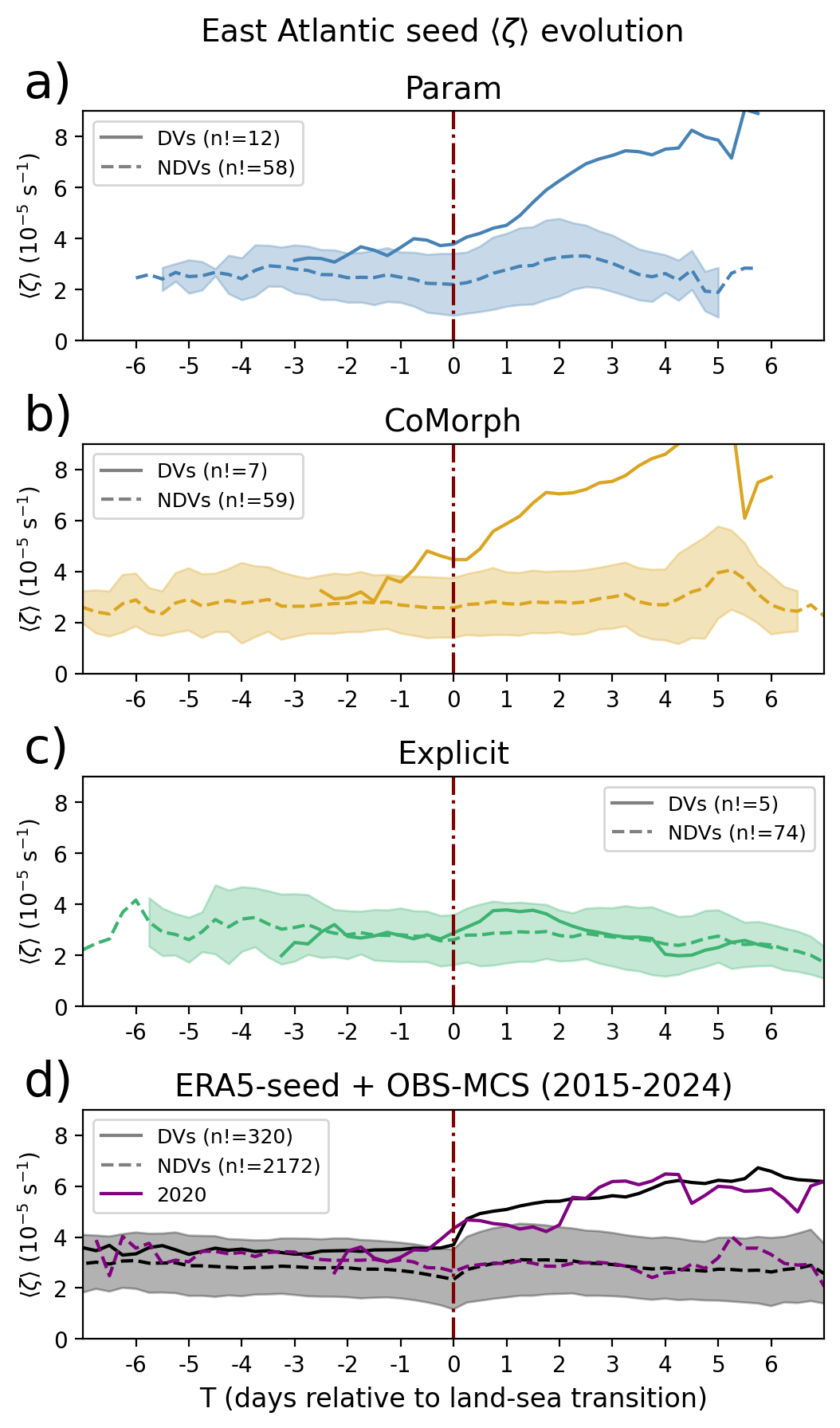}
    \includegraphics[width=0.49\textwidth]{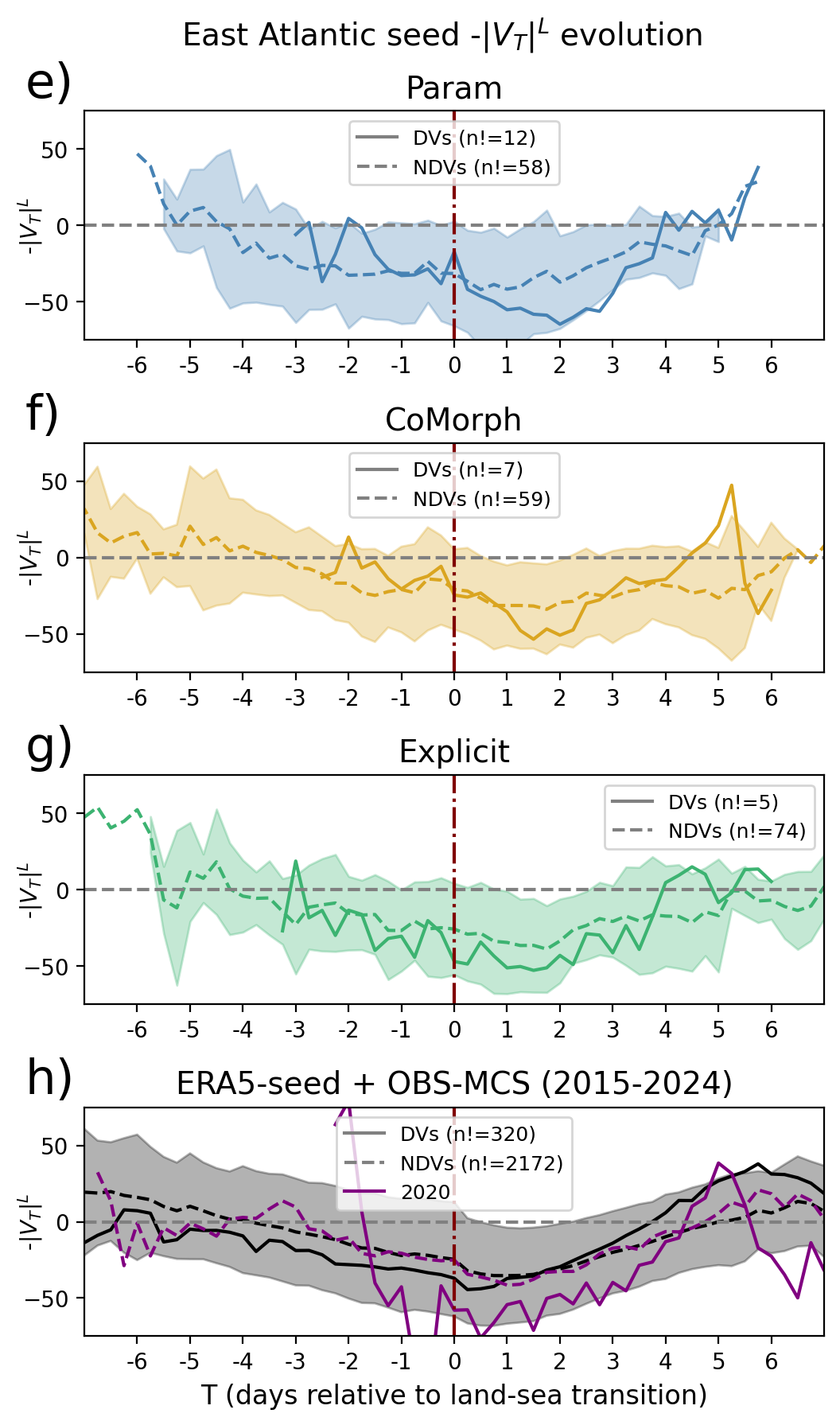}
    \caption{Composite mean evolution for developer (DV, solid) and non--developer (NDV, dashed) East Atlantic seed vortices which transition from land to ocean. Panels \textit{(a---d)} plot vertically averaged lower--tropospheric relative vorticity $\langle\zeta\rangle$ for each MetUM model and ERA5, respectively. In panel \textit{(d)} black lines and grey shading show ERA5 1979---2024 climatological mean, purple lines show 2020 means. Column repeated in panels \textit{(e---h)} for low--level thermal wind, \hartL. All composites centred on timestep $T$=0 at which vortex first detected over ocean. Shading represents NDV population standard deviation. Total DV and NDV track counts specified in legends; the number of sampled vortices varies with $T$, with some timestep counts specified in Fig.~\ref{fig:LS_profile_evolution}.}
    \label{fig:LS_vortex_evolution}
\end{figure}

Figures~\ref{fig:LS_vortex_evolution}a--d show the evolution of mean T63 resolution vertically averaged relative vorticity, $\langle\zeta\rangle$, for developer (solid lines) and non--developer vortices (dashed) which transition from West Africa to the ocean. All vortex trajectories are labelled by a time coordinate $T$ set such that $T$=0 identifies the first timestep at which the vortex centre lies over the ocean; positive $T$ values indicate later oceanic evolution. Only timesteps where vortices lie within the \eawaf domain are included in the composite means. Time--referenced results shown here are consistent with longitude--referenced composites (not shown).

The Figure shows that the distinction of DVs from the full seed vortex population depends on processes which occur around the West African coast. Results here are consistent with similar findings presented by \citeA{Duvel2021vortices} using an alternative tracking method in ERA--Interim reanalysis. In all models, we find that DV mean vorticity diverges away from the NDV population around $T$=0. However, in \ral{} (Fig.~\ref{fig:LS_vortex_evolution}c) the DV vorticity growth is very small, remaining within the NDV standard deviation range (shading). In contrast, in  \gal{}, \comorph{} and ERA5 (Figs.~\ref{fig:LS_vortex_evolution}a,b,d), DVs have already strengthened over land, such that at $T$=0 their vorticity is beyond the NDV spread. DVs then show the most rapid vorticity amplification in their first 2 days over the ocean --- far prior to eventual TCG. NDVs meanwhile are characterised by little or no vortex amplification over the ocean, remaining at the strength set by their land--based evolution. Overall, vorticity evolution in ERA5 during 2020 (\ref{fig:LS_vortex_evolution}d, purple lines) follows climatology (black) for DVs and NDVs, although amplification for DVs is weaker in their first two days offshore. In all models, there is a subtle decline in mean NDV vorticity in the days before the vortex reaches the coast. This in part reflects vortex genesis events near the coast (e.g. over the Guinea highlands, Fig.~S3a), but also points to the relatively limited influence of deep--inland AEW activity on likelihood that a seed will undergo cyclogenesis: it is behaviour around the coast which is critical.

There are also differences between the thermodynamic structures of DVs and NDVs leaving the African continent. Figures~\ref{fig:LS_vortex_evolution}e--h show that in the days following coastal transition, comparing solid versus dashed lines shows DVs typically have a stronger low--level cold core structure than NDVs, as measured by the Hart phase space parameter \hartL. This evolution appears delayed in \gal{} (\ref{fig:LS_vortex_evolution}e) and \comorph{} (\ref{fig:LS_vortex_evolution}f), versus \ral{} (\ref{fig:LS_vortex_evolution}g) and ERA5 (\ref{fig:LS_vortex_evolution}h), highlighting the pronounced role of coastal, rather than land--based, processes in the parameterised models. Previous work by \citeA{Hopsch2010analysis} further supports the hypothesis that DVs in the MetUM simulations are delayed in generating warm cores. TCs themselves are axisymmetric full--depth warm core structures, and indeed DVs are seen to develop low--level warm cores after roughly four days in ERA5 climatology, consistent with NATL TCG typically being in the western reaches of the MDR (Fig.~\ref{fig:tctracks}). Around the coast, the colder cores of DVs versus NDVs suggest they have a more pronounced mid--level vortex structure: through PV inversion, a strong mid--level circulation causes lower temperatures near the surface, and higher upper--level temperatures. In 2020, ERA5 DVs show much colder cores than the climatological mean (Fig.~\ref{fig:LS_vortex_evolution}h, purple vs black lines), suggesting this season's seeds had stronger mid--level circulations than average.  The timing of low--level DV warm core development in our results is consistent with a circulation deepening from the mid--troposphere \cite{Bister1997genesis,Raymond2014tropical}. Here all vortices show warm upper cores (positive \hartU) at the coastal transition (not shown), with higher values for DVs, consistent with \citeA{Brammer2015variability}.

\begin{figure}[h!]
    \centering
    \includegraphics[width=\textwidth]{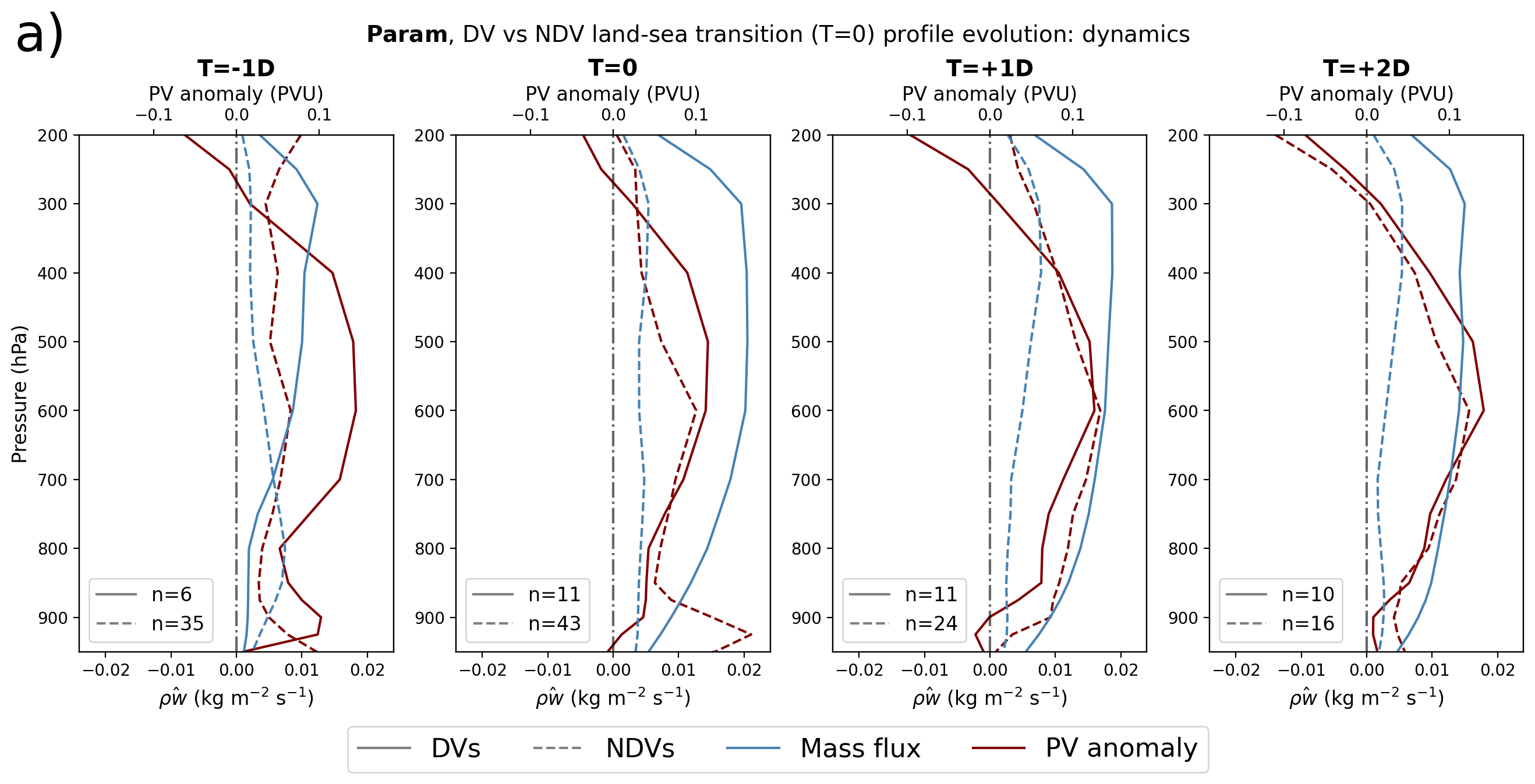}
    \includegraphics[width=\textwidth]{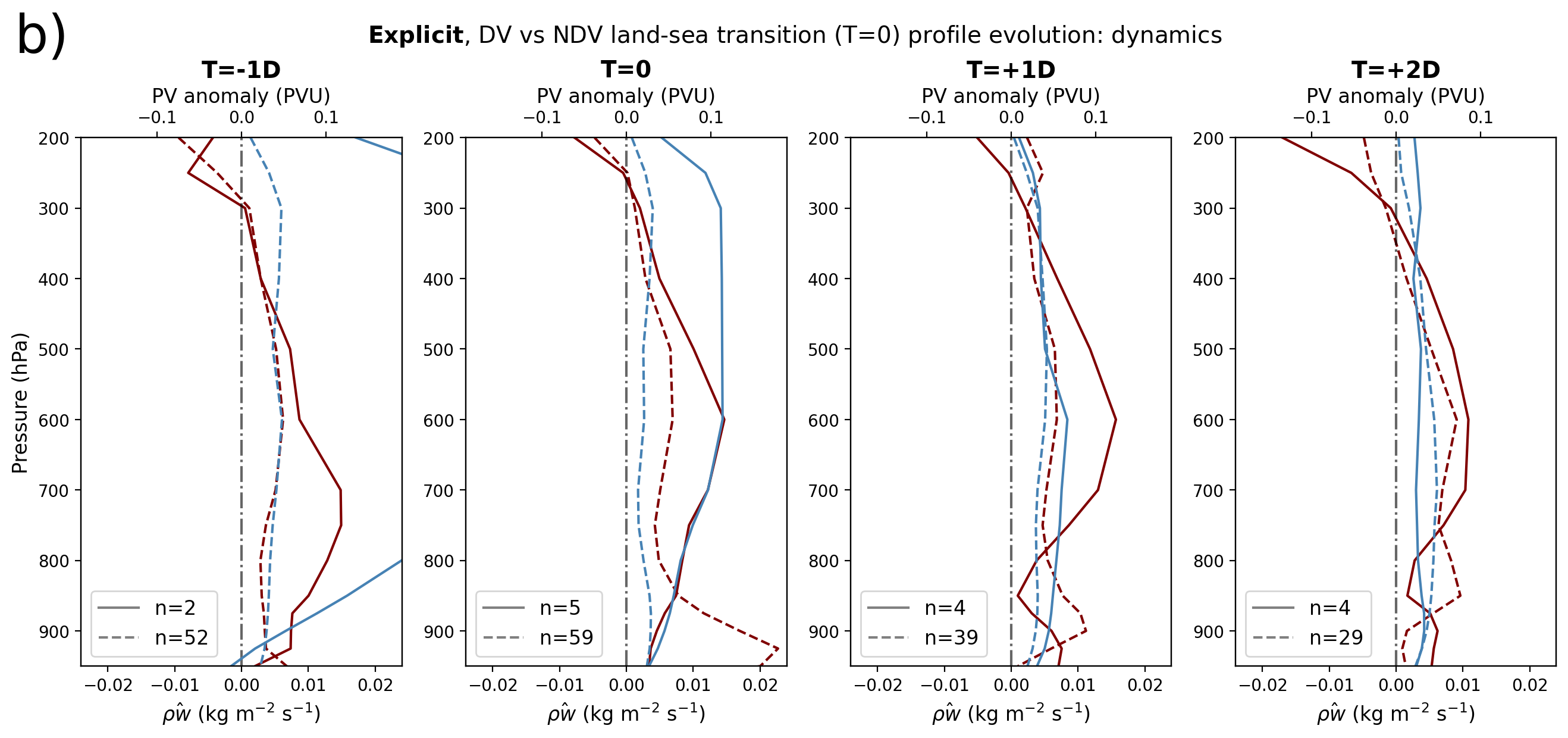}
    \caption{\textit{(a)} Composite mean evolution of vertical profiles of \gal{} vortex contour--area normalised vertical mass flux ($\hat{\rho w}$, blue) and contour--area mean PV anomaly (red), for East Atlantic DVs (solid) and NDVs (dashed) which transition from land to ocean. Time coordinate $T$ as in Fig.~\ref{fig:LS_vortex_evolution}, numbers in panel legends list number of DVs and NDVs, respectively, at each $T$. \textit{(b)} Repeated for \ral.}
    \label{fig:LS_profile_evolution}
\end{figure}

To understand how the structure of vortices in the simulated \eawaf seed populations evolves, we plot composite vertical profiles in the days around $T$=0. In \gal{} (Fig.~\ref{fig:LS_profile_evolution}a), consistent with the cold low--level cores (Fig.~\ref{fig:LS_vortex_evolution}e) we find maximum positive vortex contour--area mean PV anomalies (red) around 600\,hPa. The mid--level PV maximum for DVs (solid lines) over land is characteristic of AEWs, emphasising their role in providing seed vortices. By $T$=1D, DVs and NDVs (dashed) share comparable PV anomaly profiles below around 400\,hPa. They are instead distinguished by their relative vertical mass flux profiles $\rho w$ (blue), with consistently stronger mass fluxes about DVs. Moreover, not only is $\rho$w stronger for DVs, but near the coast ($T$=0) the profile is top--heavy. By $T$=2D the mass profiles are more balanced, pointing to the development of further bottom--heavy mass fluxes from deep convective updrafts. At all timesteps the DV profiles show ``deep--inflow'', whereby vertical mass flux increases with height \cite{Seidel2026convective}. In contrast, in NDVs a strong balanced mass flux does not evolve and there is also little vorticity amplification.

Profile evolution in \gal{} for DVs versus NDVs is consistent with that in both \comorph{} (Fig.~S\SuppRef{}) and a complementary analysis of NCAR--NCEP reanalysis by \citeA{Leppert2013relation}. Results for \ral{} (Fig.~\ref{fig:LS_profile_evolution}b), in contrast, are very different, with far weaker PV anomalies and vertical mass fluxes. \ral{} DVs do show a mid--level PV anomaly that is stronger than NDVs, which only have a low--level maximum, however there is little growth from $T$=-1D, which is perhaps to be expected since the DVs do not intensify to TC strengths until further along their track (Fig.~\ref{fig:tctracks}c). DVs over land and at the coast (note small population) have top--heavy convective profiles which are much stronger than for NDVs, as in \gal{} (Fig.~\ref{fig:LS_profile_evolution}a). However once offshore, strong top--heavy mass fluxes are not maintained over the ocean for DVs in \ral{}, nor do bottom--heavy fluxes from deep convection evolve to balance the profile as in \gal{} and \comorph{}: indeed, for \ral{} the mean 800\,hPa mass flux for NDVs is larger than that for DVs at $T$=2D.

It is well documented in the literature that vertical mass flux profiles play a key role in the amplification of seed vortices through vortex stretching \cite{Bister1997genesis,Raymond2007theory,Raymond2014tropical}. By mass conservation, a top--heavy, stratiform flux profile causes mid--level horizontal convergence, which leads to vertical stretching and mid--level circulation growth. If the stretching tendency is maintained, a low--level circulation can develop if there are also deep convective updrafts which provide bottom--heavy mass flux profiles and thus low--level horizontal convergence. Balanced, deep--inflow vertical mass flux profiles, such as those for \gal{} in Fig.~\ref{fig:LS_profile_evolution}a, enable precisely this established mechanism, and indeed a circulation budget analysis confirms that vortex stretching is significantly stronger for DVs than NDVs, with mid--level stretching strongest around the West African coast, and low--level stretching strongest in the Ocean subdomain. Full details are provided in Supporting Information (Fig.~S\SuppRef{}) 

\begin{figure}[t]
    \centering
    \includegraphics[width=\textwidth]{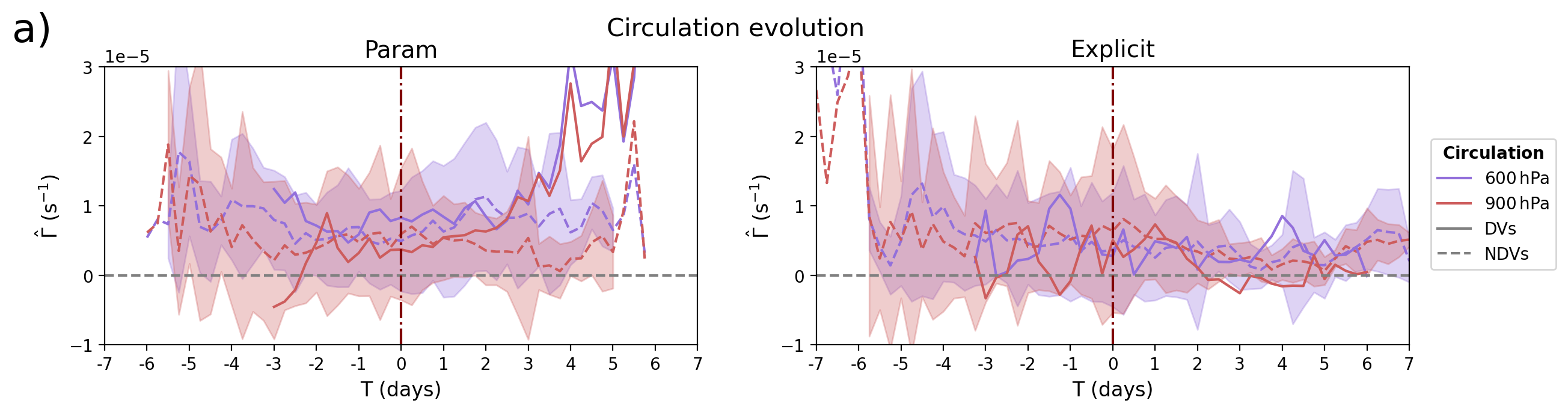}
    \includegraphics[width=\textwidth]{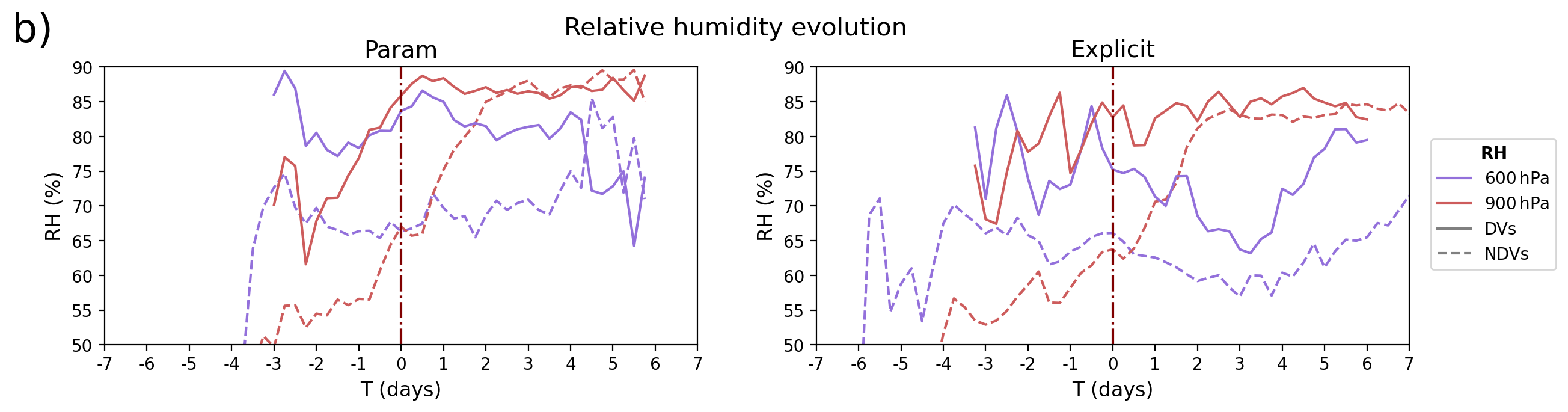}
    \includegraphics[width=\textwidth]{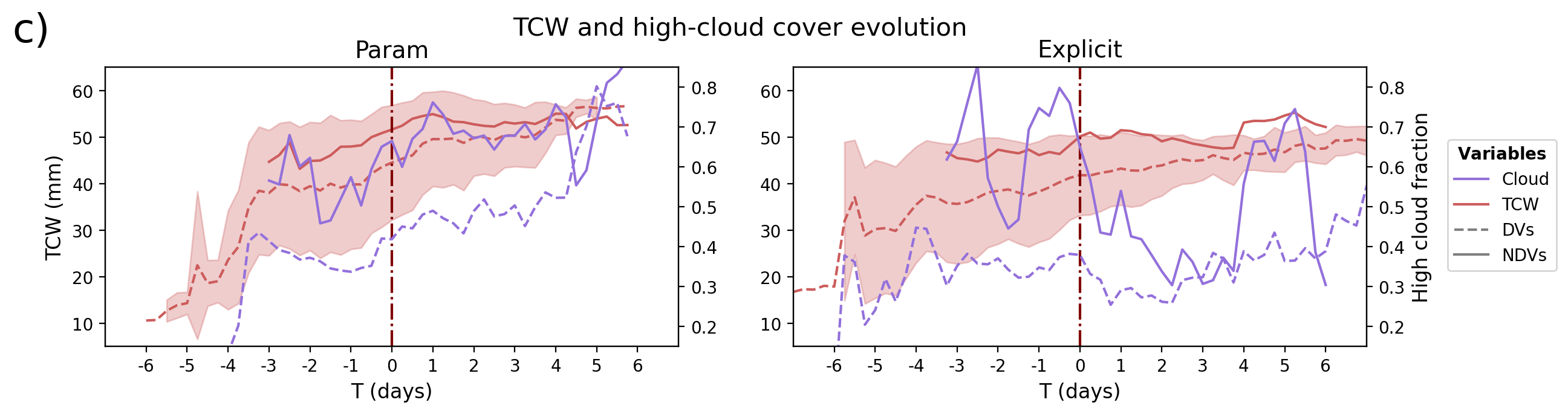}
    \caption{Full composite evolution of East Atlantic DV (solid) and NDV (dashed) vortex fields at given pressure levels in \gal{} (left column) and \ral{} (right column). \textit{(a)} Area--normalised circulation $\hat{\Gamma}$ at 600\,hPa (purple) and 900\,hPa (red). Composites centred on timestep $T$=0, shading represents standard deviation for NDV population; the total number of vortices varies with time. \textit{(b)} Repeated for area--mean relative humidity at 600\,hPa and 900\,hPa, respectively. \textit{(c)} Repeated for area--mean TCW (red) and high--cloud cover (purple). NDV spread for RH and high--cloud cover is uninformative and thus omitted for clarity.}
    \label{fig:galRAL_plev_evolution}
\end{figure}

The deepening of the precursor mid--level vortex in \gal{} DVs can be clearly seen in composite evolution plots for vortex circulation $\Gamma$ on distinct pressure levels (Fig.~\ref{fig:galRAL_plev_evolution}a). Informed by PV profiles in Fig.~\ref{fig:LS_profile_evolution}, we use 600\,hPa and 900\,hPa to pick out a mid--level maximum and strengthening near--surface vortex, respectively. 
The overall evolution is (by definition) consistent with results for $\langle\zeta\rangle$ (Fig.~\ref{fig:LS_vortex_evolution}a,c), but here the coloured profile split offers more detail.

Looking at the solid lines at $T$=0 in \gal{}, we see that for DVs the mean circulation at 600\,hPa is over twice that at 900\,hPa, confirming the existence of a distinct precursor mid--level vortex. The NDV dashed lines for the two levels are more comparable, indicating less pronounced mid--level vortices. The DV mid--level circulation (purple, solid) remains approximately constant until around $T$=2D, whereas the NDV 600\,hPa mean (purple, dashed) grows to equal that of the DV population. In contrast, during the same period 900\,hPa circulation (red) remains approximately constant for NDVs (dashed), but steadily intensifies for DVs (solid). After $T$=2D the DV profiles are balanced, and rapidly grow beyond the NDV standard deviation shown by shading. \gal{} DVs thus undergo amplification via low--level circulation development, consistent with the maintenance of strong balanced mass flux profiles. There is little correlation between deep--inland circulation values and vortex development. This also holds in \ral{} (right panel), however the lack of distinction between DV and NDV population means continues offshore, with DV values for $\Gamma$ always within the NDV shading range. For \ral{} DVs there are no signatures of a prominent mid--level circulation, since the 600\,hPa and 900\,hPa solid lines have similar magnitudes, and no growth at 900\,hPa. Indeed, the DV population 900\,hPa mean is negative after $T$=2D, suggesting frictional dissipation overcomes any positive tendencies (Fig.~S7c).

The departure of DV vertical mass fluxes from the NDV mean in \gal{} (Fig.~\ref{fig:LS_profile_evolution}a) leads that of the vortex circulation values, showing that convection instigates the split in development pathways. The strong balanced mass flux profiles in \gal{} are maintained beyond $T$=2D, and in \ral{} do not develop further offshore within the \eawaf domain (not shown). We conclude that top--heavy and balanced mass flux profiles in \gal{} reinforce mid--level circulations and drive low--level circulation development in DVs through vortex stretching that is absent from NDVs. The lack of vorticity amplification in \ral{} is due to the failure of DVs to maintain such mass flux profiles. 

Why then are strong balanced profiles maintained for \gal{} (and \comorph{}) DVs, but not \ral{}? To answer this, we first compare the evolution of relative humidity (RH) as the vortices propagate offshore (Fig.~\ref{fig:galRAL_plev_evolution}b) in the different simulations. Inland and at the coast, comparing solid and dashed lines in both panels shows DV area--mean RH is consistently higher than for NDVs. In \gal{}, 600\,hPa RH (purple) is higher than that at 900\,hPa (red) onshore and for DVs, is only 5\% lower offshore, whereas in \ral{} DVs show a significant, rapid decline in RH at 600\,hPa--only as vortices move further offshore. A similar decline is found at 800\,hPa (not shown). This mid--level drying during DV evolution after $T$=0 in \ral{} partly reflects the dry offshore bias in mean seasonal column moisture (Fig.~\ref{fig:wam_mean}g), with Fig.~\ref{fig:galRAL_plev_evolution}c showing DV--centred TCW (solid red) gradually declines after $T$=1D in \ral{}. However, domain mean environmental TCW does still increase offshore, and the NDV population means (dashed) show no similar rapid declines and behave similarly to those in \gal. Local humidity profiles vary strongly with degrees of convective organisation due to convection's control of cloud structures: our results for \ral{} suggest that the  mid--level drying in \ral{} DVs versus \gal{} is a further reflection of the differences in their vertical mass flux evolution.

Moist environments are themselves favourable for vortex development through decreased static stability \cite{Raymond2007evolution} and positive feedbacks on the strength and vertical profiles of convective mass fluxes \cite{Seidel2026convective}. The full seed population in \gal{} (and \comorph) does consistently show higher TCW values than \ral{} (Fig.~\ref{fig:galRAL_plev_evolution}c, red lines and shading), increased mean state $\theta_e$ values (Figs.~\ref{fig:wam_mean}e--g and~S\SuppRef{}), and stronger, more top--heavy vertical mass fluxes. However, we cannot causally isolate the role of moisture in driving seed evolution. Vortex amplification and convective organisation both cause mid--level moistening, making it difficult to distinguish the relative control of environmental and seed factors. Moreover, we find JJASO seasonal mean vertical mass fluxes are comparable between the simulations, and in fact strongest in \ral{} (Fig.~S8), including at upper levels. This further points to a larger role for local vortex processes, rather than larger--scale feedbacks, in driving the differences in seed evolution. 

In that regard, \citeA{Luschen2026stratiform} have recently shown that stratiform and anvil cloud radiative forcing enhance the vertical mass flux and mid--level humidity about seed vortices by suppressing stratiform downdrafts. The key role of cloud populations in driving vortex evolution within our MetUM simulations is confirmed by the evolution of mean high--cloud areal coverage over vortex footprints (Fig.~\ref{fig:galRAL_plev_evolution}c, purple): for coastal DVs in \gal{} (solid, left panel) the mean high--cloud fraction is around 0.7, whereas in \ral{} (right panel) the fraction plummets after $T$=0 to below 0.4. Comparing solid and dashed purple lines, we see the mean fraction is consistently higher about DVs than NDVs in both simulations, but always larger around vortices in \gal{}. The changes in cloud fraction are consistent with results from observed DVs \cite{Leppert2013relation} and the mechanisms whereby environmental humidity increases convective mass flux, since \citeA{Seidel2026convective} identified a correlation with convective area fraction rather than updraft intensity. We therefore turn our focus to interactions of convective clouds with seed vortices.

\subsection{Role of Mesoscale Convective Systems}
\label{sec:mcs}

Deep convective activity and high anvil cloud cover over our East Atlantic study region is primarily organised into MCSs \cite{Janiga2014convection,Wu2024contribution}, strongly motivating an examination of the behaviour and physics of MCS activity in the MetUM simulations.

\begin{figure}[t]
    \centering
    \includegraphics[width=\textwidth]{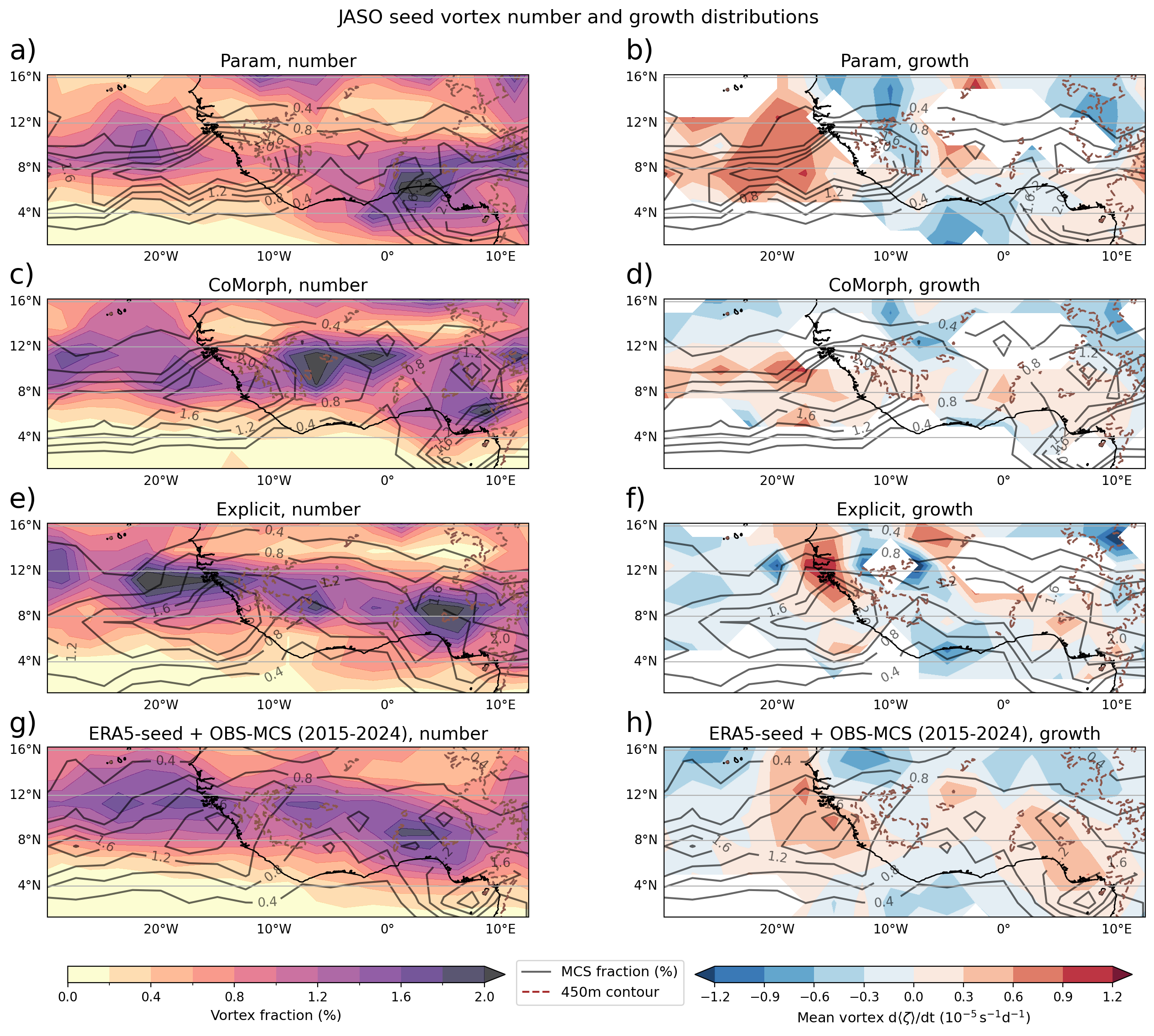}
    \caption{Fractional number distributions (LH column, shading) and mean 6--hourly $\langle\zeta\rangle$ tendencies (RH column, shading) for JASO seed vortices. Solid contours in both columns show MCS fractional number distribution, with all spatial distributions derived from 2.5$^\circ$ bins. Rows show results for (\textit{a,b}) \gal{}; (\textit{c,d}) \comorph{}; (\textit{e,f}) \ral{}; and (\textit{g,h}) ERA5 seed vortex and OBS--MCS 2015---2024 means. Dashed brown contours show 450\,m orography contour; white space in RH column indicates masking of means underpinned by less than five values. All fractional distributions are taken with respect to the displayed domain.}
    \label{fig:vortex_spatial_distributions}
\end{figure}

Figure~\ref{fig:vortex_spatial_distributions} shows that in satellite observations and all MetUM simulations, the MCS number distribution peaks around the Guinea coast. This is a well known result driven by the orographic influence of the Guinea highlands (e.g. \citeA{Hodges1997distribution,Wu2024contribution}), with the high number of MCSs causing an offshore regional rainfall maximum (Fig.~\ref{fig:wam_mean}d). We restrict to JASO to focus on the established monsoon phase during which MCSs are most active, and use the 2015--2024 MCS climatology period as observational benchmark, with results for 2020 very similar (Fig.~S\SuppRef{}a).

In \gal{}, the coastal MCS peak (contours) is centred approximately 2$^\circ$S of that in other datasets, but this peak is collocated with the primary seed vortex train (Fig.~\ref{fig:vortex_spatial_distributions}a, shading), reflecting the southern--shifted AEJ in this simulation. Figure~\ref{fig:vortex_spatial_distributions}b shows the corresponding spatial map of vorticity amplification over the region. A maximum in \gal{} vortex amplification occurs several degrees west of the MCS peak, pointing to the potential causal role of the MCS population in driving the amplification through their balanced convective and stratiform flux profiles. Offshore peak $\dd\langle\zeta\rangle/\dd t$ values are collocated with the vortex train, and will be dominated by the DV subset (Fig.~\ref{fig:LS_vortex_evolution}a). There is little robust vorticity growth inland that is removed from orography (dashed contours).

These qualitative patterns are very similar in \comorph{} (Fig.~\ref{fig:vortex_spatial_distributions}c,d) and the ERA5 climatology (Fig.~\ref{fig:vortex_spatial_distributions}g,h), except that there are more widespread regions of weak vorticity growth inland, particularly across the flat Sahel (e.g. 0--10$^\circ$W, 7--12$^\circ$N): here, \gal{} shows a dry rainfall bias, and thus will have weaker diabatic heating, limiting vortex amplification. The southern confinement of Sahel MCSs in parameterised models is a known bias explained by a limited response to environmental vertical wind shear \cite{Maybee2024wind}, but one that clearly has little impact on the promotion of DVs given the simulations' final TC counts. We find that continental seed vorticity growth does still correlate with later cyclogenesis: the spatial mean Land $\dd\langle\zeta\rangle/\dd t$ value for ERA5 DVs is 4.3$\times$10$^{-6}$\,s$^{-1}$d$^{-1}$, beyond the interquartile range of all Land ERA5 seeds (Fig.~S\SuppRef{}). The same result holds in JRA--3Q and over the Coastal subdomain for both reanalyses, but there with greater absolute magnitude, with Coastal DV mean ERA5 $\dd\langle\zeta\rangle/\dd t$ = 9.0$\times$10$^{-6}$\,s$^{-1}$d$^{-1}$.

The spatial growth distribution for \ral{} (Fig.~\ref{fig:vortex_spatial_distributions}f) is confined to the coast, and supports our previous time--composite analysis showing vortex amplification is weak compared to the other simulations. Comparing against the number distribution (\ref{fig:vortex_spatial_distributions}e) shows much of the high mean growth region is caused by relatively few seeds: there is only limited overlap with the primary seed train, with the peak offshore vortex density (11$^\circ$N, 21$^\circ$W) corresponding to a negative mean vorticity tendency. This is not the case in the other simulations or reanalysis. In \ral{}, the coastal MCS peak lies 2.5$^\circ$S of the offshore seed vortex train (Fig.~\ref{fig:vortex_spatial_distributions}e, contours vs shading). There is no such offset in \gal{} (\ref{fig:vortex_spatial_distributions}a) and \comorph{} (\ref{fig:vortex_spatial_distributions}c), but a small southward displacement between peak MCS and seed number is apparent between OBS--MCS and ERA5--seed number peaks (\ref{fig:vortex_spatial_distributions}g; holds for both climatology and 2020, Fig.~S9a). A latitudinal offset in--itself between MCS and vortex trains therefore does not prevent vortex development.

Indeed despite the differences in systems' spatial alignment, the much larger absolute MCS population in \ral{} (317 coastal MCSs) versus the parameterised simulations (104 in \gal) ensures the likelihood of finding an MCS neighbouring a seed vortex is equivalent. For example, the probability of a Coastal vortex lying within 2$^\circ$ of an MCS centroid is 0.45 in \ral{} and 0.38 in \gal{} (Fig.~S9b). These probabilities increase for seeds which undergo vorticity growth in \ral{} in particular, pointing to the importance of MCS interactions for its small DV population. However, comparing shading and contours down the right column of Fig.~\ref{fig:vortex_spatial_distributions} shows \ral{} is the only dataset for which there is no direct overlap between peak MCS activity and mean offshore vorticity amplification. This suggests that, on average, the MCSs which do interact with the seed vortex population in \ral{} must suffer unphysical characteristics that contribute to the lack of offshore vorticity growth evident in Fig.~\ref{fig:vortex_spatial_distributions}f. Conversely, the much tighter proximity of MCS and vortex peaks in \gal{} (\ref{fig:vortex_spatial_distributions}a) than climatology (\ref{fig:vortex_spatial_distributions}g) likely contributes to a mean Coastal seed vortex growth rate larger than the reanalysis medians (Figs.~\ref{fig:vortex_spatial_distributions}b and~S10).

\begin{figure}[t]
    \centering
    \includegraphics[width=\textwidth]{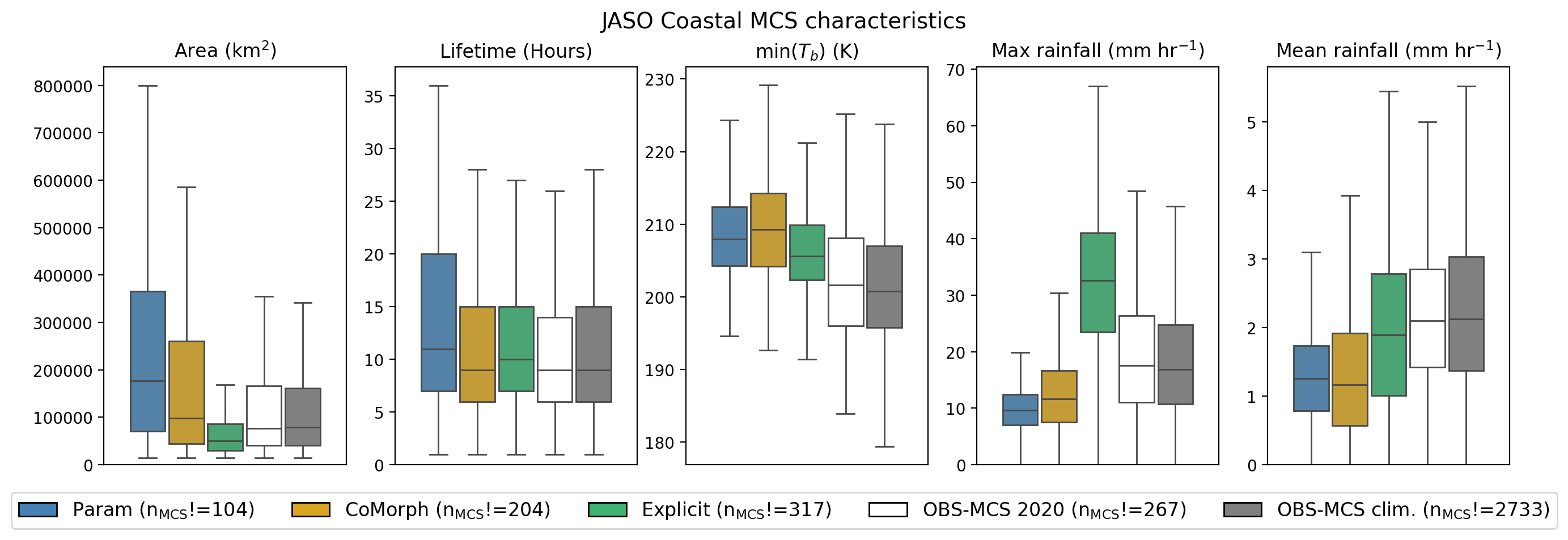}
    \caption{Distributions of system area, lifetime, minimum $T_b$, maximum and mean rainfall, for Coastal subdomain MCSs. Distributions taken over all individual hourly snapshots for all measures except lifetime, with outlier values excluded. Whisker lengths are 1.5$\times$ the interquartile range. All outputs derived from common 0.1$^\circ$ grid; the Coastal subdomain is 10$^\circ$--20$^\circ$W, 5$^\circ$--15$^\circ$N.}
    \label{fig:mcs_boxplots}
\end{figure}

Figure~\ref{fig:mcs_boxplots} shows that there are systematic differences in the structures and biases of tracked Coastal MCS in the three simulations. MCSs in \gal{} and \comorph{} have significantly larger areas and reduced convective intensities (as measured by minimum system $T_b$ and rainfall maxima) compared to \ral{} and observed counterparts. \gal{} in particular shows a significant tail of extremely large MCS anvils ($>$400,000km$^2$) and extremely long--lived systems ($>$1 day), echoing similar results for MCSs in the GAL9 configuration when run at higher resolution \cite{Maybee2024wind}. In contrast, anvils in \ral{} are far smaller than in OBS--MCS, with the third--quartile comparable to the observed median in both 2020 and climatology. Storms in \ral{} then show significantly overestimated rainfall maxima versus OBS--MCS (rainfall from IMERGv7). The interquartile ranges for rainfall maxima in \gal{} and \comorph{} are more realistic but biased too low, and their storms underestimate the high--rainfall skew of the observed distribution. For all variables, 2020 seasonal OBS--MCS characteristics are very similar to the climatology.

For MCS mean rainfall, which combines storm snapshot structure and intensity, \ral{} exhibits a more realistic distribution than \gal{} and \comorph{}, reflecting these simulations' systematic structural biases of too--large MCSs with too--low convective intensities. Yet storms in \ral{} also suffer a structural bias, of too--small MCSs with too--high convective intensities. The proclivity for smaller, more intense systems in \ral{} is associated with more bottom--heavy mass flux profiles, and indeed likely explains the increased offshore mean low--level regional mass fluxes (Fig.~S8c) found relative to \gal{} and \comorph{}, since there are more, stronger deep convective towers.

\begin{figure}[t]
    \includegraphics[width=\textwidth]{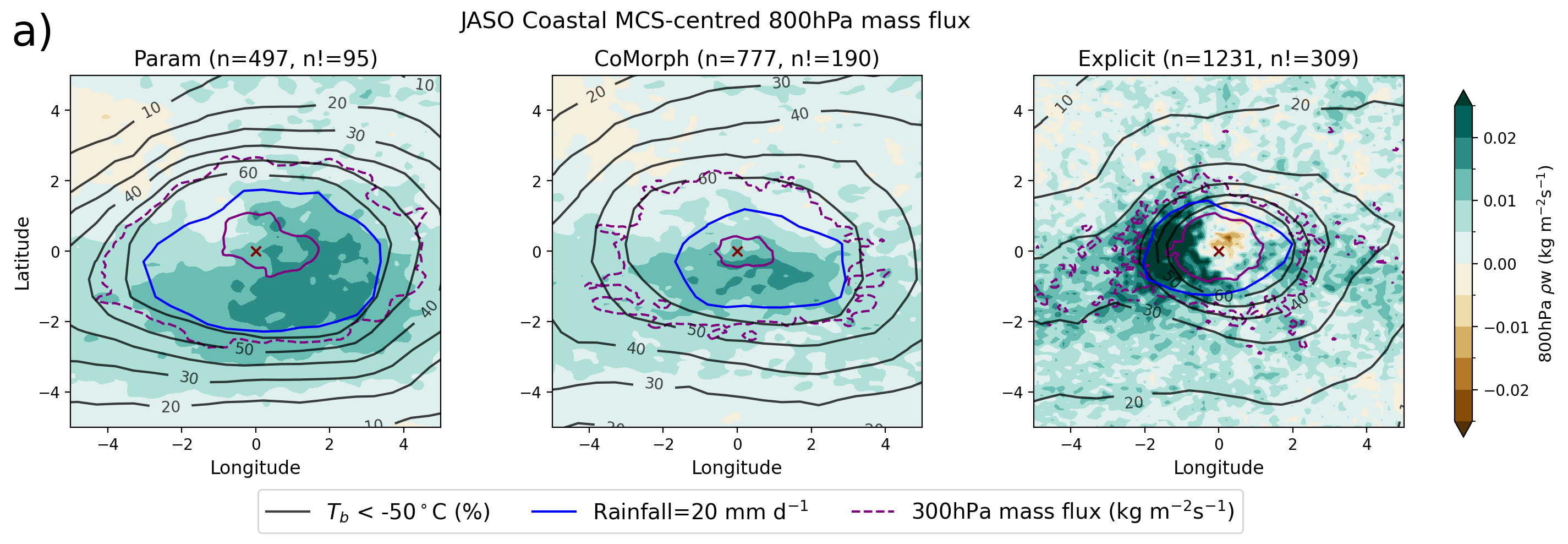}
    \includegraphics[width=\textwidth]{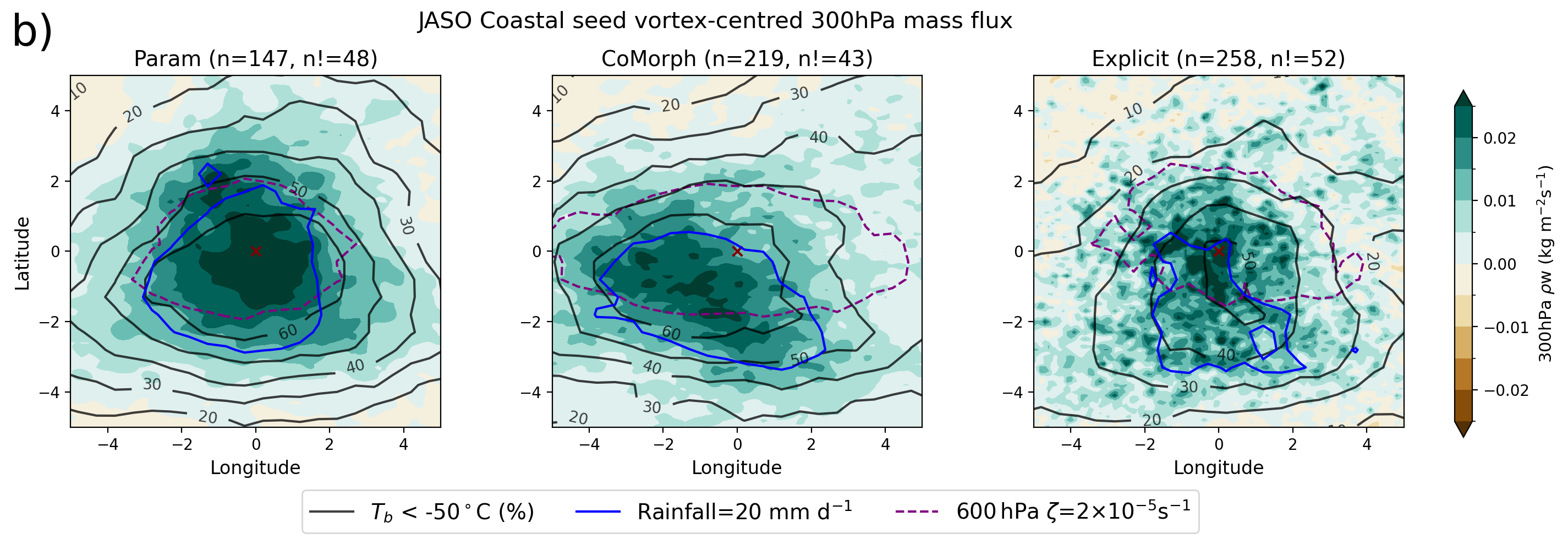}
    \caption{\textit{(a)} JASO mean Coastal MCS min($T_b$)--centred composites of mean 0.1$^\circ$ regridded 800\,hPa  vertical mass flux (shading) and cold cloud occurrence ($T_b$$<$-50$^\circ$C, grey contours). Purple contours show regions with equivalent composite mean 300\,hPa mass flux exceeding 0.01\,kg\,m$^{-2}$s$^{-1}$ (dashed) and 0.04\,kg\,m$^{-2}$s$^{-1}$ (solid). Solid blue contour demarks 20 mm day$^{-1}$ rainfall. Values in title parentheses specify total number of sampled MCS snapshots ($n$) and the number of unique systems ($n!$). Values are lower than in Fig.~\ref{fig:mcs_boxplots} due to 3--hourly frequency of vertical profile diagnostics. \textit{(b)} Repeated for 300\,hPa mass fluxes centred on JASO Coastal seed vortex tracked centres, with purple dashed contour showing 600\,hPa 2$\times10^{-5}$ s$^{-1}$ relative vorticity ($\zeta$). Here title values giving respective vortex counts for 6--hourly sampling (TRACK frequency).} 
    \label{fig:smorgasbord}
\end{figure}

These differences in MCS structure are key for the offshore seed populations since they affect the structure of the vertical mass flux profiles generated by those systems. To examine MCS--seed interactions, in Fig.~\ref{fig:smorgasbord} we compare composited mass fluxes and convective activity in both Coastal MCS and seed vortices (rows) and in each MetUM simulation (columns). Grey contours show the probability of cold cloud cover ($T_b<$-50$^\circ$C), while a single blue contour highlights high rainfall regions (20 mm day$^{-1}$).

From the contours, we see that all MCS composites (Fig.~\ref{fig:smorgasbord}a) show a core central region of strong 300\,hPa vertical mass flux (solid purple, $>$0.04\,kg\,m$^{-2}$s$^{-1}$; shading in Fig.~S\SuppRef{}a) and high rainfall rates; such attributes are to be expected from compositing on organised convective systems. More revealing are differences in updraft structure visible from 800\,hPa mass fluxes (shading) and the cold cloud probability contours. \gal{} and \comorph{} show uniformly positive and weak low--level MCS mass flux, with no apparent structure. In contrast, in \ral{} there is a region of intense positive mass flux west of the composite centre, with strong negative mass flux at the centre, and a heterogeneous distribution of much weaker fluxes towards the composite edges. \ral{} thus captures the observed structure of West African squall lines, with strong inflow into MCS updrafts leading a rainfall--evaporation driven downdraft that generates surface cold pools (e.g.~\citeA{Maybee2025how}). Crucially, the homogenous low--level structure in the parameterised models does not match reality. In \ral{}, the 50\% cold--cloud occurrence contour is tightly aligned with the central convective region that exhibits a mesoscale circulation, whereas in \gal{} and \comorph{} it spreads over 1$^\circ$ further from the composite centre, reflecting the large--anvil skew in these models' tracked MCS area distributions. This skew extends to the MCS stratiform mass fluxes: at 300\,hPa in the parameterised models, persistent weak positive fluxes (dashed purple contour) continue to extend beneath the larger anvil shields, with homogenous positive values $>$0.01 kg\,m$^{-2}$s$^{-1}$ extending towards the longitudinal composite edges. The correlation with anvil extent also holds in \ral{}, where the equivalent contour is constrained to the composite centre and shows a far more heterogeneous, intermittent structure.

Turning to the vortex--centred composites (Fig.~\ref{fig:smorgasbord}b), the qualitative patterns of convective fields mirror those for MCSs. In particular, \gal{} and \comorph{} show larger regions of homogeneously uniform 300\,hPa stratiform mass flux, whereas in \ral{} the field is far less coherent. Intense convective mass fluxes from MCS cores do not superpose to give a strong, widespread mean signal as in \gal{}, and the smaller MCS anvil sizes and deficient stratiform mass flux contributions inherently give rise to a more heterogeneous mean picture, which is also found at 800\,hPa (Fig.~S11b). In \gal{} and \comorph{}, although the strongest mass fluxes are weaker, the significantly larger MCS stratiform decks lead to the smoother, more uniform fluxes experienced by seeds, perhaps due to the deep convection scheme providing a more linear response to environmental variables than the explicit convection in \ral{}. The key difference in MCS anvil areas is again shown by the grey cold cloud probability contours, with the neighbourhood of a parameterised seed vortex far more likely to be under high, cold cloud cover than in \ral{}. Moreover, in \gal{} the mid--level vortex (purple dashed contour), high rainfall (blue contour), most frequent convective activity (grey contours) and strongest mass fluxes (shading) all overlap and are centred on the seeds. In \ral{}, both convective cloud and rainfall peak due south of the vortex centre, with the most organised high--rainfall region 2$^\circ$ away, outside the 600\,hPa vorticity contour.

We thus arrive at an explanation for the reduced mass flux profiles available to seeds in \ral{} versus \gal{} (and \comorph), and thereby the reduced number of East Atlantic seeds which develop into TCs. The two simulations show systematic differences in MCS structure: systems in \ral{} suffer a small--anvil bias, with upper--level vertical mass fluxes concentrated into more vigorous, distinct updrafts, compared to unrealistically large storms in \gal{} that have a weaker, more homogeneous updraft structure covering a much broader area, which aggregates to give a strong top--heavy mean mass flux. While the probability of a vortex lying near MCS convection is higher in \ral{}, that of a vortex lying under an MCS anvil, and thus being influenced by a consistent deep--inflow mass flux profile, is higher in \gal{} due to systems' larger extent (Fig.~\ref{fig:smorgasbord}b). Moreover, the primary vortex and MCS trains are latitudinally aligned (Fig.~\ref{fig:vortex_spatial_distributions}), and the large MCSs in these trains persist longer than \ral{} equivalents, enabling longer--lasting influences on vortex development. Our finding that the extents and latitudinal positions of MCSs are their main features modulating tropical cyclogenesis from AEW seeds, rather than absolute intensities, agrees with observational analysis \cite{Leppert2013relation,Nunez2020wave}. Finally, background environmental conditions in \gal{} and \comorph{} are more favourable for vortex development, with moister atmospheric columns and higher low--level temperatures ($T$ and $\theta_e$) than both ERA5 and \ral{} (Fig.~\ref{fig:wam_mean}) contributing towards more favourable conditions for sustaining convection. These key simulation differences, and the key processes controlling DV evolution, are summarised in Fig.~\ref{fig:schematic}.


\section{Discussion}
\label{sec:discuss}


\begin{figure}[t]
    \centering
    \includegraphics[width=\textwidth]{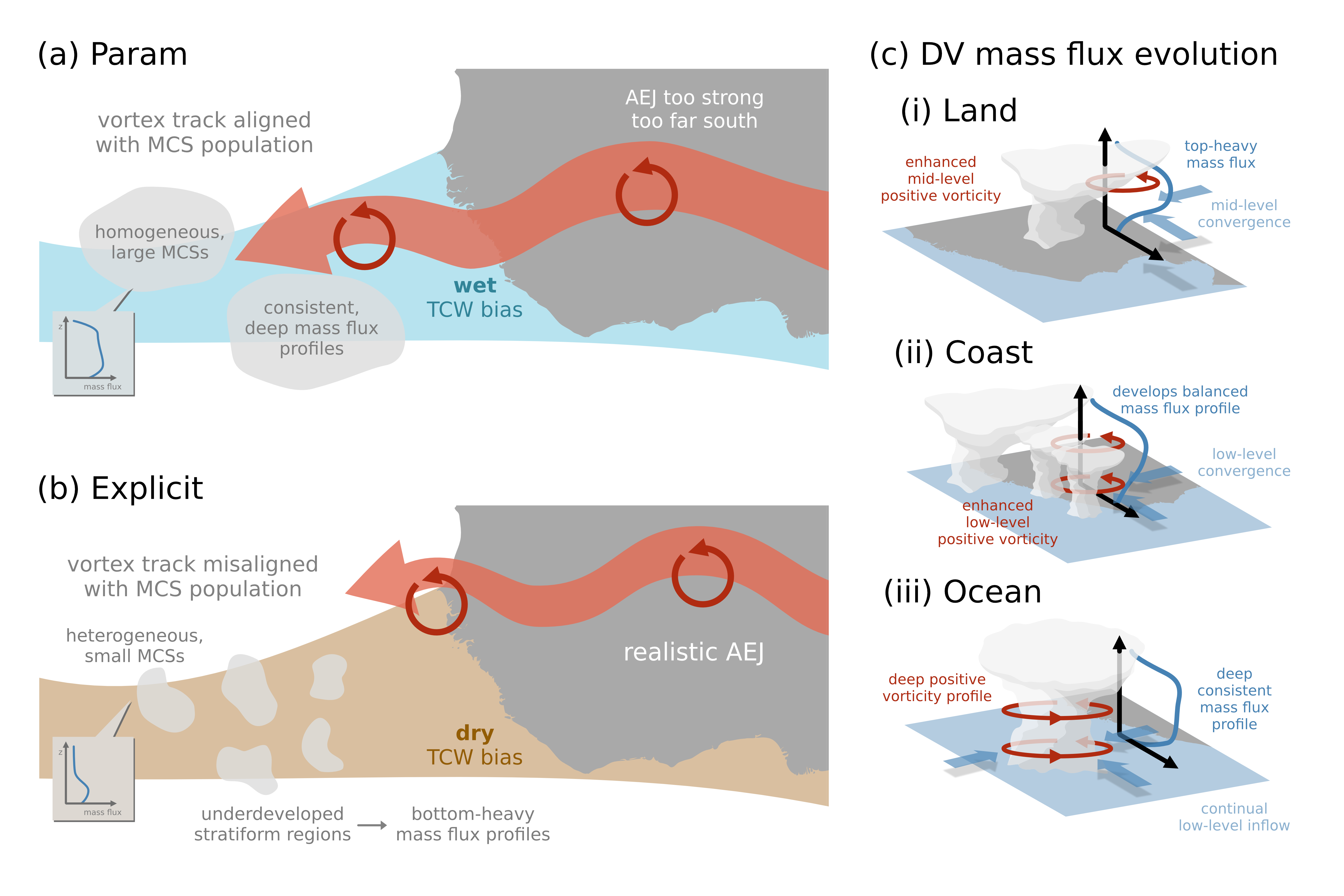}
    \caption{Schematics of key processes and simulation differences in seed vortex evolution described in this paper. In panels \textit{(a)} and \textit{(b)}, shading over ocean represents large--scale bias in TCW (versus ERA5), while the continental broad wavy arrows (red shading) represent the relative position and strength of model AEJs. Red circular arrows embedded in the AEJ denote seed vortex trains, and grey shaded objects represent MCS population characteristics. Grey insets give representative local mass flux profiles experienced by seed vortices. Panel \textit{(c)} summarises how vertical mass flux profiles (blue) from MCSs (translucent light grey clouds) around developer vortices (DVs) contribute towards mid and low--level circulation growth (red circular arrows) as DVs propagate offshore. Blue horizontal arrows represent convergence/vortex stretching. Objects are not to scale.}
    \label{fig:schematic}
\end{figure}

Our results show that over one NATL hurricane season, a high--resolution global CP simulation has much weaker TC activity than coarser--resolution, parameterised counterparts driven by the same SSTs and generating similar \eawaf seed vortex frequencies. We have argued this is due to differences in seed vortex development, linked to simulation biases in convective organisation and MCS structure. In drawing conclusions from our results, it is critically important to consider the cascading role of compound biases. The very small numbers of \ral{} TC developers originating from West Africa is not a result of the simulation being unable to simulate the physical mechanisms through which vortex development occurs in this pathway: in Sec.~\ref{sec:vortices}, we saw DVs do have different characteristics to NDVs, with stronger top--heavy mass flux profiles, more pronounced mid--level PV anomalies, moister mid--levels and enhanced vortex stretching. The issue is that on average, far fewer East Atlantic seed vortices receive the essential ingredients for undergoing development. Offshore seeds in \ral{} suffer from a drier, cooler atmosphere than those in reanalysis and the parameterised models (Figs.~\ref{fig:schematic}a,b), which is an essential ingredient for developing the convective mass flux signatures required for cyclogenesis (Fig.~\ref{fig:schematic}c). While \ral{} simulates perhaps the most realistic Coastal MCSs in terms of storms' mesoscale structure and mean rainfall, their convective intensities are too strong. Stratiform rainfall and anvil extents are also underestimated, yielding far weaker and less coherent top--heavy vertical mass fluxes near vortices than in parameterised simulations (Fig.~\ref{fig:smorgasbord}). Over land, all results indicate more active convection and vortex amplification in \ral{}, consistent with previous findings regarding the benefit of convection permitting simulations for AEWs and the monsoon circulation \cite{Marsham2013role,Tomassini2017interaction,Russell2020african}. While results from reanalysis affirm the importance of land--based seed characteristics and interactions for DVs, they also highlight the even stronger role of coastal processes. 

The far more active vortex--development pathway, and thus NATL hurricane population, found here in \gal{} does not show a ``superior'' simulation in the sense of physical realism. None of the MetUM simulations were intended as faithful reflections of the observed 2020 hurricane season. \gal{} may best reflect the season's extremely strong activity, but it does so for several wrong reasons which together interfere constructively when it comes to \eawaf seed vortices (Fig.~\ref{fig:schematic}a). Firstly, the mean West African monsoon state is far from reanalysis and observations, with an AEJ that is too strong, too far south, and extends too far offshore. Atmospheric columns are moister and warmer than reanalysis, and onshore, rain is confined too far south versus IMERGv7. These mean state biases affect vortex development two--fold: the AEJ characteristics push the vortex trains leaving West Africa further south than climatology, and provide an uncommonly favourable background environment for coastal MCSs and vortices. The shift in the vortex train causes direct spatial alignment with peak MCS activity, thereby leading to an increased propensity for the mechanisms summarised in Fig.~\ref{fig:schematic}c to occur. We know from Fig.~\ref{fig:mcs_boxplots} that the simulated MCSs are themselves misrepresentative of real systems, yet the extremely broad anvils with homogeneous mean upper--level vertical mass fluxes they generate remain favourable for vortex development \cite{Leppert2013relation,Luschen2026stratiform}. The critical nature of offshore vorticity growth for TC development ensures that more destructive onshore biases in \gal{} play a far more minor role.

Results from \comorph{} show that a modern experimental convective parameterisation is able to reduce these biases in the continental monsoon while maintaining favourable conditions offshore for vortex development. Analysis of \comorph{} corroborates the vortex amplification pathway we have traced in \gal{} (Fig.~S6 and Fig.~\ref{fig:schematic}c), but again this occurs as a result of constructive interference of simulation biases, with unphysical, homogeneous MCS structures in particular. In contrast, for \ral{} the model biases interfere destructively (Fig.~\ref{fig:schematic}b): seed vortices propagate offshore north of peak MCS activity into a lower--troposphere that is too dry. The intense, overly--convective MCSs that vortices do interact with then offer unfavourable vertical mass flux profiles for vortex amplification versus parameterised simulations. The existence of (smaller) latitudinal offsets between peak offshore ERA5--seed and OBS--MCS activity suggest that it is biases in MCS stratiform rainfall which are key. This is a systematic issue that continues to afflict most km--scale CP models \cite{Feng2025mesoscale}. The root cause remains unknown, but overestimated convective updraft widths caused by underestimated entrainment, which then bias key microphysics processes, are a candidate explanation \cite{Lebo2015effects,Prein2021sensitivity,Kukulies2024simulating}. However, the configuration used in \ral{} has demonstrated significant improvement in the handling of entrainment processes in continental West African MCSs \cite{Maybee2024wind}: further research is urgently required to solve this widespread modelling problem. 

Turning from a model--development to a physical perspective, it is interesting to consider the role of fundamental vortex dynamics in influencing the different simulated seed vortex populations. All our results pertain to vortices identified by the TRACK algorithm from T63 spectrally--truncated vorticity fields. This spectral smoothing, while enabling fair model comparison and filtering out small--scale vorticity noise, also hides the native--grid eddies which ultimately make up the mesoscale seed vortices of interest. At the convective storm scale, diabatic heating generates horizontal dipoles which can extend through the troposphere \cite{Lilly1986structure,Chagnon2009horizontal}. While individual PV anomalies are strong, the dipole structure averages to a relatively weaker composite vortex. The detailed structure of sub--MCS--scale dipoles is part--resolved in \ral{} ($\sim$5\,km/n2560 resolution), but unresolved in \gal{} and \comorph{} ($\sim$10\,km/n1280), which instead generate larger, smoother coherent vorticity fields (Fig.~S\SuppRef{}). In all simulations, there is a significant correlation ($p<$0.05) between TRACK vortex maximum and mean area, and system lifetime. The smoother vorticity dynamics in the parameterised models may thus artificially extend the persistence of the larger mesoscale vortices, potentially providing another compounding factor that favours DVs in \gal{} and \comorph{}. The complex, highly non--linear dynamics of vortex--vortex interactions can lead to upscale energy transfer and thus perturbation growth in (midlatitude) circulations \cite{Oertel2021quantifying}. Large--domain CP models, particularly at lower grid spacings than \ral{}, offer an unprecedented opportunity to study the details of such interactions in realistic settings, and should certainly be pursued in the context of TC development.

Our results also offer insights for ongoing research surrounding the relative role of seed populations and large--scale environmental conditions in controlling tropical cyclogenesis \cite{Emanuel2022tropical,Wang2025definition}. Many authors have used statistical genesis indices based on background environmental fields to explain trends in TC variability \cite{Camargo2007use,Tippett2011poisson}, while spontaneous cyclogenesis has been described in model experiments \cite{Nolan2007tropical,Wing2016role}. The potentially limited role of the base--state \eawaf seed population in NATL tropical cyclogenesis has also been demonstrated through numerous experiments suppressing AEW activity and finding limited changes, or even increases, in TC populations \cite{Patricola2018response,Danso2022influence,Bercos2023effects,Kouski2026influence}. In our study, mean state fields alone are not enough to explain inter--model TC variability: all show favourable Atlantic basin conditions for cyclogenesis (Fig.~S2). The seed frequency also offers little explanatory power. In the framework of seed--based genesis indices \cite{Vecchi2019tropical,Hsieh2020large,Emanuel2022tropical}, we find the key factor is the transition probability of a seed vortex to a TC, which is far lower in \ral{}. Issues with this factor have been identified globally in other CP models \cite{Emlaw2026understanding}. This perspective also perhaps offers a connection to AEW--suppression experiments, in that suppressing the seed frequency will have limited effect on cyclogenesis rates if the transition probabilities are high. Ensembles of such experiments have found increased East Atlantic (our Ocean subdomain) humidity and rainfall \cite{Kouski2026influence}, implying increased convective activity and thereby increased seed--TC transition probabilities in a similar manner to the constructive action of biases in \gal{}. We leave detailed investigation of the intersection with such experimental frameworks for future work.


\section{Conclusions}
\label{sec:conclusions}

Interactions between organised deep convection and mesoscale seed vortices in the Eastern Atlantic play a fundamental role in the lifecycle of Atlantic TCs. Here we have explored these interactions and their relationship with TC variability in three global, year--long Met Office Unified Model simulations, reanalysis data, and satellite observations. All simulations are high resolution and driven by observed (2020) SSTs. We find that a $\sim$5\,km grid, convection permitting model (\ral) produces almost half of the number of NATL basin TCs as equivalent $\sim$10\,km configurations with different convection parameterisation schemes (\gal{} and \comorph), despite all simulations having equivalent East Atlantic seed population sizes and characteristics (Fig.~\ref{fig:aej_seeds_overview}). The zonal AEW--pathway to NATL cyclogenesis is most strongly affected (Fig.~\ref{fig:tctracks}), and simulation differences cannot be explained by regional environmental conditions in the central and western Atlantic, which are favourable for cyclogenesis in all three simulations (Fig.~S2).

We show that different rates of NATL TC development from the East Atlantic seed vortex population stem from the dynamics of vortices as they transition from continental Africa to the Atlantic ocean. In \gal{}, \comorph{}, and ERA5 climatology, such developer vortices (DVs) show strong vorticity tendencies offshore that are absent in \ral{}. This difference can be explained by the relative interaction of MCSs with seed vortices, which we and many previous authors argue act as a key catalyst for vortex development (e.g. ~\citeA{Schwendike2010convection,Russell2020african,Nunez2020wave}). Mean state biases in the West African monsoon in \gal{} act favourably to latitudinally align the primary vortex train with strong MCS activity and high atmospheric column moisture (Fig.~\ref{fig:schematic}a). Systematic differences in MCS structure stemming from the representation of deep convection lead to unrealistically large, long--lived anvils accompanied by broader regions of positive upper--level mass flux in \gal{} and \comorph{} versus \ral{} (Fig.~\ref{fig:smorgasbord}). Parameterised storms are characterised by very large regions of uniform weak intensity, in contrast to those in \ral{} and observations which show heterogeneous distributions of smaller storms with more intense convection (Figs.~\ref{fig:mcs_boxplots} and~\ref{fig:smorgasbord}a). MCSs in \ral{} have a bias towards small anvil sizes and overestimated convective rainfall, in common with other km--scale CP models \cite{Feng2025mesoscale}.

In this paper we have identified an important upscale consequence of this long--standing model bias. Through analysis of composite vortex profiles, we demonstrate that differences between DVs and NDVs, and thus cyclogenesis rates, stem primarily from seeds' vertical mass flux profiles (Fig.~\ref{fig:LS_profile_evolution}). These are stronger and more balanced for DVs, enabling cyclogenesis via low--level circulation growth through maintenance and stretching of an existing mid--level circulation (\citeA{Bister1997genesis,Raymond2007theory,Raymond2014tropical}; Fig.~\ref{fig:schematic}c). The key MCS feedback here is the essential provision of the offshore deep--inflow, stratiform dominated mass flux. The unrealistic homogeneous structures of parameterised MCSs provide this, whereas the overly--convective storms in \ral{} with underestimated anvils and stratiform rainfall do not  (Figs.~\ref{fig:mcs_boxplots} and~\ref{fig:smorgasbord}), leading to weak basin--wide TC activity. Our results reinforce the finding of \citeA{Leppert2013relation} that convective area coverage is more important for TC development from AEWs than convective intensity, and are consistent with recent analysis showing the positive feedback of anvils' radiative forcings on vortex amplification \cite{Luschen2026stratiform}. In future work we will evaluate the detailed cloud processes contributing to MCS simulation biases in full detail. Here, we have focussed on the role of local, mesoscale processes in driving vortex evolution in the simulations. An important extension would be to consider the role of known synoptic drivers of AEWs and cyclogenesis, for example equatorial waves \cite{Lawton2022influence,Feng2023equatorial,Rios2024modulation}. The details of convective organisation cause equatorial wave biases in operational forecasts that are improved by some CP climate models \cite{Gehne2022diagnostics,Takasuka2026precipitation}, pointing to an important overlapping topic for future research, and potential opportunity for shorter--scale operational km--scale forecasting systems.

A major limitation of this study is the lack of observations for investigating the vertical structures of real MCS and seed vortex populations. While West African coastal processes arguably remain under--explored versus those over the continental Sahel, extensive field campaigns and modelling efforts are rapidly improving our knowledge of this key region \cite{Nowottnick2024dust,Flamant2024cyclogenesis,Sakaeda2025synoptic,Martinez2025evaluating,Colon2026convective}. ERA5 cannot be considered observational truth for mesoscale dynamics, as exemplified by its own small 2020 NATL TC count (Fig.~\ref{fig:tctracks}d). We have relied on inference from ERA5 vortices and satellite--observation MCS tracks to make comments about which MetUM simulation offers the most realistic seed population dynamics and vortex structures, but cannot confirm the precise causal mechanisms for real systems beyond known case studies (e.g.~\citeA{Bister1997genesis,Schwendike2010convection,Raymond2011vorticity}). In this context we look forward to investigating the potential step--change offered by new satellite missions measuring vertical soundings (EarthCARE, \citeA{Illingworth2015earthcare}; INCUS, \citeA{Dolan2023time}), and their ability to test the mechanisms outlined here in the real atmosphere (e.g.~\citeA{Roh2025vertical}).

Finally, our results contribute to the ongoing development of global CP climate models. West Africa is a natural testbed for such simulations due to the critical upscale role of deep convection in regional circulation, and previous studies already highlight the benefits of explicit convection in this region \cite{Marsham2013role,Pante2019resolving}. Simulation of TCs, cyclogenesis, AEWs and MCSs all individually benefit from explicit convection. Analysis of \ral{} shows the details of their interactions remain difficult, and that downstream or upscale improvements are not guaranteed. The present simulations are too short to isolate whether this is simply natural variability, or a more systematic bias. Robust intercomparison of MCS tracks from existing DYAMOND models shows systematic underestimation of their stratiform component \cite{Feng2025mesoscale}, while results from \citeA{Emlaw2026understanding} and TC tracking activities at the 2025 WCRP Global km--scale hackathon suggest that the underestimation of NATL hurricane activity in global CP models is not unique to the MetUM, and afflicts all available DYAMOND3 simulations. Our future work will explore whether these problems are related. Existing global parameterised configurations, such as GAL9 in \gal{}, benefit from decades of tuning and operational testing. The next generation of global models will have to follow a similar path before achieving their potential benefits.


\section*{Acknowledgements}

This work was primarily funded through the NERC Hurac\'{a}n grant NE/W009587/1. FM is funded by First Rains: UKRI Future Leaders Fellowship MR/W011379/1. BM is grateful to Xiangbo Feng for providing ERA5 reanalysis data and insightful comments; and to Alex Baker, Sally Lavender, John Marsham, Doug Parker, Michael Reeder, Lorenzo Tomassini, and the Lawrence Berkeley Lab CASCADE project team, for conversations which shaped the development of the article. The clarity and scientific presentation of our paper benefited enormously from insightful comments by three anonymous reviewers. We thank the organisers of the 2025 WCRP global km--scale modelling hackathon under the Digital Earth Lighthouse for catalysing the development of the MetUM models studied, and for fostering a collaborative community. We acknowledge use of the huracanpy, tcpyPI, metpy, scipy and xarray python pacakges, and heartily thank their developers. Our work relied on JASMIN, the UK national collaborative data analysis facility.

\section*{Open Research}

Code and data underpinning this research are available publicly at \citeA{Maybee2026zenodo}, which includes seed vortex and MCS tracking data in readily accessible formats. 2D fields from original versions of the MetUM simulations are available from the Digital Earths Global Hackathon Data Catalogue at https://digital-earths-global-hackathon.github.io/catalog/. All satellite observation products and reanalysis fields used in this research are available publicly.

\section*{Supporting Information}

Supporting Information can be found with the online version of this article.

\section*{Conflicts of Interest}
The authors declare there are no conflicts of interest for this manuscript.

\bibliography{references}

@article{Pante2019resolving,
	title={{Resolving Sahelian thunderstorms improves mid-latitude weather forecasts}},
	author={Pante, Gregor and Knippertz, Peter},
	journal={Nature Communications},
	volume={10},
	number={1},
	pages={3487},
	year={2019},
	doi={https://doi.org/10.1038/s41467-019-11081-4}
}

@article{Tomassini2017interaction,
	title={{The interaction between moist diabatic processes and the atmospheric circulation in African Easterly Wave propagation}},
	author={Tomassini, Lorenzo and Parker, Douglas J and Stirling, Alison and Bain, Caroline and Senior, Catherine and Milton, Sean},
	journal={Quarterly Journal of the Royal Meteorological Society},
	volume={143},
	number={709},
	pages={3207--3227},
	year={2017},
	doi={https://doi.org/10.1002/qj.3173}
}

@article{Russell2020potential,
  title={The potential vorticity structure and dynamics of {A}frican easterly waves},
  author={Russell, James OH and Aiyyer, Anantha},
  journal={Journal of the Atmospheric Sciences},
  volume={77},
  number={3},
  pages={871--890},
  year={2020},
  doi={https://doi.org/10.1175/JAS-D-19-0019.1}
}

@article{Russell2020african,
  title={{African Easterly Wave dynamics in convection-permitting simulations: Rotational stratiform instability as a conceptual model}},
  author={Russell, James OH and Aiyyer, Anantha and Dylan White, J},
  journal={Journal of Advances in Modeling Earth Systems},
  volume={12},
  number={1},
  pages={e2019MS001706},
  year={2020},
  doi={https://doi.org/10.1029/2019MS001706}
}

@article{Maybee2024wind,
	title={{Wind shear Effects in Convection--Permitting Models Influence MCS Rainfall and Forcing of Tropical Circulation}},
	author={Maybee, Ben and Marsham, John H and Klein, Cornelia and Parker, Douglas J and Barton, Emma Jane and Taylor, Christopher M and Lewis, Huw and Sanchez, Claudio and Jones, Richard Wilson and Warner, James Lewis},
	journal={Geophysical Research Letters},
	volume={51},
	number={17},
	year={2024},
	doi={https://doi.org/10.1029/2024GL110119}
}

@article{Maybee2025how,
	title={{How sensitive are Sahelian Mesoscale Convective Systems to cold pool suppression?}},
	author={Maybee, B and Bassford, J and Marsham, J H and Lewis, H and Field, Paul and Klein, C and Parker, D J},
	journal={Quarterly Journal of the Royal Meteorological Society},
    volume={151},
    number={772},
	year={2025},
	doi={https://doi.org/10.1002/qj.5032}
}

@article{Stein2014three,
	title={The three-dimensional morphology of simulated and observed convective storms over southern {E}ngland},
	author={Stein, Thorwald HM and Hogan, Robin J and Hanley, Kirsty E and Nicol, John C and Lean, Humphrey W and Plant, Robert S and Clark, Peter A and Halliwell, Carol E},
	journal={Monthly Weather Review},
	volume={142},
	number={9},
	pages={3264--3283},
	year={2014},
	doi={https://doi.org/10.1175/MWR-D-13-00372.1}
}

@article{Emanuel2022tropical,
  title={{Tropical Cyclone Seeds, Transition Probabilities, and Genesis}},
  author={Emanuel, Kerry},
  journal={Journal of Climate},
  volume={35},
  number={11},
  pages={3557--3566},
  year={2022},
  doi={https://doi.org/10.1175/JCLI-D-21-0922.1}
}

@article{Marsham2013role,
	title={{The role of moist convection in the West African monsoon system: Insights from continental--scale convection--permitting simulations}},
	author={Marsham, John H and Dixon, Nick S and Garcia-Carreras, Luis and Lister, Grenville MS and Parker, Douglas J and Knippertz, Peter and Birch, Cathryn E},
	journal={Geophysical Research Letters},
	volume={40},
	number={9},
	pages={1843--1849},
	year={2013},
	doi={https://doi.org/10.1002/grl.50347}
}

@article{Birch2014seamless,
	title={{A seamless assessment of the role of convection in the water cycle of the West African Monsoon}},
	author={Birch, Cathryn E and Parker, DJ and Marsham, JH and Copsey, D and Garcia-Carreras, L},
	journal={Journal of Geophysical Research: Atmospheres},
	volume={119},
	number={6},
	pages={2890--2912},
	year={2014},
	doi={https://doi.org/10.1002/2013JD020887}
}

@article{Raymond2014tropical,
  title={Tropical cyclogenesis and mid-level vorticity},
  author={Raymond, David J and Gjorgjievska, Sa{\v{s}}ka and Sessions, Sharon and Fuchs, {\v{Z}}eljka},
  journal={Australian Meteorological and Oceanographic Journal},
  volume={64},
  number={1},
  pages={11--25},
  year={2014},
  doi={https://doi.org/10.1071/ES14003}
}

@article{Nunez2020wave,
  title={{A wave-relative framework analysis of AEW--MCS interactions leading to tropical cyclogenesis}},
  author={N{\'u}{\~n}ez Ocasio, Kelly M and Evans, Jenni L and Young, George S},
  journal={Monthly Weather Review},
  volume={148},
  number={11},
  pages={4657--4671},
  year={2020},
  doi={https://doi.org/10.1175/MWR-D-20-0152.1}
}

@article{Schwendike2010convection,
  title={{Convection in an African easterly wave over West Africa and the eastern Atlantic: A model case study of Helene (2006)}},
  author={Schwendike, Juliane and Jones, Sarah C},
  journal={Quarterly Journal of the Royal Meteorological Society},
  volume={136},
  number={S1},
  pages={364--396},
  year={2010},
  doi={https://doi.org/10.1002/qj.566}
}

@article{Morris2024synoptic,
  title={{How is synoptic-scale circulation influenced by the dynamics of mesoscale convection in convection-permitting simulations over West Africa?}},
  author={Morris, Fran and Schwendike, Juliane and Parker, Douglas J and Bain, Caroline},
  journal={Journal of the Atmospheric Sciences},
  volume={81},
  number={4},
  pages={765--782},
  year={2024},
  doi={https://doi.org/10.1175/JAS-D-22-0032.1}
}

@article{Morris2025closing,
  title={Closing the circulation budget},
  author={Morris, F and Robinson, CM and Reeder, M and Schwendike, J and Parker, DJ and Bain, CL and Short, CJ},
  journal={Journal of Geophysical Research: Atmospheres},
  volume={130},
  number={2},
  pages={e2024JD041738},
  year={2025},
  doi={https://doi.org/10.1029/2024JD041738}
}

@article{Raymond2011vorticity,
  title={The vorticity budget of developing typhoon {N}uri (2008)},
  author={Raymond, D J and L{\'o}pez Carrillo, C},
  journal={Atmospheric Chemistry and Physics},
  volume={11},
  number={1},
  pages={147--163},
  year={2011},
  doi={https://doi.org/10.5194/acp-11-147-2011}
}

@article{Brammer2015variability,
  title={{Variability and Evolution of African Easterly Wave Structures and their Relationship with Tropical Cyclogenesis over the Eastern Atlantic}},
  author={Brammer, Alan and Thorncroft, Chris D},
  journal={Monthly Weather Review},
  volume={143},
  number={12},
  pages={4975--4995},
  year={2015},
  doi={https://doi.org/10.1175/MWR-D-15-0106.1}
}

@article{Mapes1995diabatic,
  title={Diabatic divergence profiles in western {P}acific mesoscale convective systems},
  author={Mapes, Brian E and Houze Jr, Robert A},
  journal={Journal of Atmospheric Sciences},
  volume={52},
  number={10},
  pages={1807--1828},
  year={1995},
  doi={https://doi.org/10.1175/1520-0469(1995)052<1807:DDPIWP>2.0.CO;2}
}

@article{Russell2017revisiting,
  title={{Revisiting the connection between African easterly waves and Atlantic tropical cyclogenesis}},
  author={Russell, James O and Aiyyer, Anantha and White, Joshua D and Hannah, Walter},
  journal={Geophysical Research Letters},
  volume={44},
  number={1},
  pages={587--595},
  year={2017},
  doi={https://doi.org/10.1002/2016GL071236}
}

@article{Wang2025definition,
  title={On the definition and tracking of tropical cyclone seeds from a climate perspective},
  author={Wang, Zhuo and Rios-Berrios, Rosimar and Stern, Daniel P and Baker, Alexander J and Beucler, Tom and Camargo, Suzana J and Duvel, Jean-Philippe and Feng, Xiangbo and Lee, Chia-Ying and Leroux, Marie-Dominique and others},
  journal={Bulletin of the American Meteorological Society},
  volume={106},
  number={9},
  pages={E1815--E1822},
  year={2025},
  doi={https://doi.org/10.1175/BAMS-D-24-0200.1}
}

@article{Kouski2026influence,
  title={{The influence of African easterly waves on Atlantic tropical cyclone tracks and landfall in large ensembles}},
  author={Kouski Jr, Ronald H and Patricola--DiRosario, Christina M and Bercos--Hickey, Emily and Risser, Mark D},
  journal={Journal of Geophysical Research: Atmospheres},
  volume={131},
  number={1},
  pages={e2025JD044501},
  year={2026},
  doi={https://doi.org/10.1029/2025JD044501}}

@article{Patricola2018response,
  title={{The response of Atlantic tropical cyclones to suppression of African easterly waves}},
  author={Patricola, Christina M and Saravanan, R and Chang, Ping},
  journal={Geophysical Research Letters},
  volume={45},
  number={1},
  pages={471--479},
  year={2018},
  doi={https://doi.org/10.1002/2017GL076081}
}

@article{Danso2022influence,
  title={{Influence of African easterly wave suppression on Atlantic tropical cyclone activity in a convection-permitting model}},
  author={Danso, Derrick K and Patricola, Christina M and Bercos--Hickey, Emily},
  journal={Geophysical Research Letters},
  volume={49},
  number={22},
  pages={e2022GL100590},
  year={2022},
  doi={https://doi.org/10.1029/2022GL100590}
}

@article{Nowottnick2024dust,
  title={{Dust, convection, winds, and waves: The 2022 NASA CPEX-CV campaign}},
  author={Nowottnick, Edward P and Rowe, Angela K and Nehrir, Amin R and Zawislak, Jonathan A and Pi{\~n}a, Aaron J and McCarty, Will and Barton-Grimley, Rory A and Bedka, Kristopher M and Bennett, J Ryan and Brammer, Alan and others},
  journal={Bulletin of the American Meteorological Society},
  volume={105},
  number={11},
  pages={E2097--E2125},
  year={2024},
  doi={https://doi.org/10.1175/BAMS-D-23-0201.1}
}

@article{Sakaeda2025synoptic,
  title={{Synoptic Modulation of the West African Coastal Atmosphere and Mesoscale Convective Systems}},
  author={Sakaeda, Naoko and Wu, Shun-Nan and Rios-Berrios, Rosimar and Martin, Elinor and N{\'u}{\~n}ez Ocasio, Kelly M and Bedka, Kristopher M and Hollis, Margaret and Lambrigtsen, Bjorn and Lawton, Quinton A and Nehrir, Amin and others},
  journal={Monthly Weather Review},
  volume={153},
  number={10},
  pages={1939--1957},
  year={2025},
  doi={https://doi.org/10.1175/MWR-D-24-0281.1}
}

@article{Takasuka2024protocol,
  title={A protocol and analysis of year-long simulations of global storm-resolving models and beyond},
  author={Takasuka, Daisuke and Satoh, Masaki and Miyakawa, Tomoki and Kodama, Chihiro and Klocke, Daniel and Stevens, Bjorn and Vidale, Pier Luigi and Terai, Christopher R},
  journal={Progress in Earth and Planetary Science},
  volume={11},
  number={1},
  pages={66},
  year={2024},
  doi={https://doi.org/10.1186/s40645-024-00668-1}
}

@article{Judt2021tropical,
  title={Tropical cyclones in global storm-resolving models},
  author={Judt, Falko and Klocke, Daniel and Rios-Berrios, Rosimar and Vanniere, Benoit and Ziemen, Florian and Auger, Ludovic and Biercamp, Joachim and Bretherton, Christopher and Chen, Xi and Dueben, Peter and others},
  journal={Journal of the Meteorological Society of Japan. Ser. II},
  volume={99},
  number={3},
  pages={579--602},
  year={2021},
  doi={https://doi.org/10.2151/jmsj.2021-029}
}

@article{Hodges2017well,
  title={How well are tropical cyclones represented in reanalysis datasets?},
  author={Hodges, Kevin and Cobb, Alison and Vidale, Pier Luigi},
  journal={Journal of Climate},
  volume={30},
  number={14},
  pages={5243--5264},
  year={2017},
  doi={https://doi.org/10.1175/JCLI-D-16-0557.1}
}

@article{Hodges1994general,
  title={A general method for tracking analysis and its application to meteorological data},
  author={Hodges, Kevin},
  journal={Monthly Weather Review},
  volume={122},
  number={11},
  pages={2573--2586},
  year={1994},
  doi={https://doi.org/10.1175/1520-0493(1994)122,2573:
AGMFTA.2.0.CO;2}
}

@article{Hodges1995feature,
  title={Feature tracking on the unit-sphere},
  author={Hodges, K},
  journal={Monthly Weather Review},
  volume={123},
  number={12},
  pages={3458--3465},
  year={1995},
  doi={https://doi.org/10.1175/1520-0493(1995)123,3458:
FTOTUS.2.0.CO;2}
}

@article{Thorncroft2001african,
  title={{African easterly wave variability and its relationship to Atlantic tropical cyclone activity}},
  author={Thorncroft, Chris and Hodges, Kevin},
  journal={Journal of Climate},
  volume={14},
  number={6},
  pages={1166--1179},
  year={2001},
  doi={https://doi.org/10.1175/1520-0442(2001)014<1166:AEWVAI>2.0.CO;2}
}

@article{Wilson1999microphysically,
  title={{A microphysically based precipitation scheme for the UK Meteorological Office Unified Model}},
  author={Wilson, Damian R and Ballard, Susan P},
  journal={Quarterly Journal of the Royal Meteorological Society},
  volume={125},
  number={557},
  pages={1607--1636},
  year={1999},
  doi={https://doi.org/10.1002/qj.49712555707}
}

@article{Wilson2008pc2,
  title={{PC2: A prognostic cloud fraction and condensation scheme. I: Scheme description}},
  author={Wilson, Damian R and Bushell, Andrew C and Kerr-Munslow, Amanda M and Price, Jeremy D and Morcrette, Cyril J},
  journal={Quarterly Journal of the Royal Meteorological Society},
  volume={134},
  number={637},
  pages={2093--2107},
  year={2008},
  doi={https://doi.org/10.1002/qj.333}
}

@article{Best2011joint,
  title={{The Joint UK Land Environment Simulator (JULES), model description--Part 1: energy and water fluxes}},
  author={Best, Martin J and Pryor, M and Clark, D B and Rooney, Gabriel G and Essery, R and M{\'e}nard, C B and Edwards, J M and Hendry, M A and Porson, A and Gedney, N and others},
  journal={Geoscientific Model Development},
  volume={4},
  number={3},
  pages={677--699},
  year={2011},
  doi={https://doi.org/10.5194/gmd-4-677-2011}
}

@article{Huffman2023integrated,
  title={{GPM IMERG Final Precipitation L3 Half Hourly 0.1 degree × 0.1 degree V07}},
  author={Huffman, George J and Stocker, EF and Bolvin, David T and Nelkin, Eric J and Tan, Jackson},
  journal={Earth Sciences Data and Information Services Center (GES DISC)},
  location={Greenbelt, MD},
  volumn={7},
  year={2023},
  doi={https://doi.org/10.5067/GPM/IMERG/3B‐HH/07}
}

@article{Hersbach2020era5,
  title={{The ERA5 global reanalysis}},
  author={Hersbach, Hans and Bell, Bill and Berrisford, Paul and Hirahara, Shoji and Hor{\'a}nyi, Andr{\'a}s and Mu{\~n}oz-Sabater, Joaqu{\'\i}n and Nicolas, Julien and Peubey, Carole and Radu, Raluca and Schepers, Dinand and others},
  journal={Quarterly Journal of the Royal Meteorological Society},
  volume={146},
  number={730},
  pages={1999--2049},
  year={2020},
  doi={https://doi.org/10.1002/qj.3803}
}

@article{Bush2025third,
	title={{The third Met Office Unified Model--JULES Regional Atmosphere and Land Configuration, RAL3}},
	author={Mike Bush and David L. A. Flack and Huw W. Lewis and Sylvia I. Bohnenstengel and Chris J. Short and Charmaine Franklin and Adrian P. Lock and Martin Best and Paul Field and Anne McCabe and Kwinten Van Weverberg and Segolene Berthou and Ian Boutle and Jennifer K. Brooke and Seb Cole and Shaun Cooper and Gareth Dow and John Edwards and Anke Finnenkoetter and Kalli Furtado and Kate Halladay and Kirsty Hanley and Margaret A. Hendry and Adrian Hill and A. Jayakumar and Richard W. Jones and Humphrey Lean and Joshua C. K. Lee and Andy Malcolm and Marion Mittermaier and Saji Mohandas and Stuart Moore and Cyril Morcrette and Rachel North and Aurore Porson and Susan Rennie and Nigel Roberts and Belinda Roux and Claudio Sanchez and Chun-Hsu Su and Simon Tucker and Simon Vosper and David Walters and James Warner and Stuart Webster and Mark Weeks and Jonathan Wilkinson and Michael Whitall and Keith D. Williams and Hugh Zhang},
	journal={Geoscientific Model Development},
	year={2025},
    volume={18},
    issue={12},
    pages={3819--3855},
	doi={https://doi.org/10.5194/gmd-18-3819-2025}
}

@article{Brown2012unified,
	title={Unified modeling and prediction of weather and climate: {A} 25--year journey},
	author={Brown, Andrew and Milton, Sean and Cullen, Mike and Golding, Brian and Mitchell, John and Shelly, Ann},
	journal={Bulletin of the American Meteorological Society},
	volume={93},
	number={12},
	pages={1865--1877},
	year={2012},
	doi={https://doi.org/10.1175/BAMS-D-12-00018.1}
}

@article{Stevens2019dyamond,
  title={{DYAMOND: the DYnamics of the Atmospheric general circulation Modeled On Non-hydrostatic Domains}},
  author={Stevens, Bjorn and Satoh, Masaki and Auger, Ludovic and Biercamp, Joachim and Bretherton, Christopher S and Chen, Xi and D{\"u}ben, Peter and Judt, Falko and Khairoutdinov, Marat and Klocke, Daniel and others},
  journal={Progress in Earth and Planetary Science},
  volume={6},
  number={1},
  pages={1--17},
  year={2019},
  doi={https://doi.org/10.1186/s40645-019-0304-z}}

@article{Field2023implementation,
  title={{Implementation of a double moment cloud microphysics scheme in the UK Met Office regional numerical weather prediction model}},
  author={Field, Paul R and Hill, Adrian and Shipway, Ben and Furtado, Kalli and Wilkinson, Jonathan and Miltenberger, Annette and Gordon, Hamish and Grosvenor, Daniel P and Stevens, Robin and Van Weverberg, Kwinten},
  journal={Quarterly Journal of the Royal Meteorological Society},
  year={2023},
  doi={https://doi.org/10.1002/qj.4414}
}

@article{vanWeverberg2021bimodal,
  title={{A bimodal diagnostic cloud fraction parameterization. Part I: Motivating analysis and scheme description}},
  author={Van Weverberg, Kwinten and Morcrette, Cyril J and Boutle, Ian and Furtado, Kalli and Field, Paul R},
  journal={Monthly Weather Review},
  volume={149},
  number={3},
  pages={841--857},
  year={2021},
  doi={https://doi.org/10.1175/MWR-D-20-0224.1}
}

@article{Lock2024performance,
  title={{The performance of the CoMorph-A convection package in global simulations with the Met Office Unified Model}},
  author={Lock, AP and Whitall, M and Stirling, AJ and Williams, KD and Lavender, SL and Morcrette, C and Matsubayashi, K and Field, PR and Martin, G and Willett, M and others},
  journal={Quarterly Journal of the Royal Meteorological Society},
  volume={150},
  number={763},
  pages={3527--3543},
  year={2024},
  doi={ https://doi.org/10.1002/qj.4781}
}

@article{Hart2003cyclone,
  title={A cyclone phase space derived from thermal wind and thermal asymmetry},
  author={Hart, Robert E},
  journal={Monthly Weather Review},
  volume={131},
  number={4},
  pages={585--616},
  year={2003},
  doi={https://doi.org/10.1175/1520-0493(2003)131<0585:ACPSDF>2.0.CO;2}
}

@article{Bercos2020relationship,
  title={{On the relationship between the African Easterly Jet, Saharan mineral dust aerosols, and West African precipitation}},
  author={Bercos--Hickey, Emily and Nathan, Terrence R and Chen, Shu-Hua},
  journal={Journal of Climate},
  volume={33},
  number={9},
  pages={3533--3546},
  year={2020},
  doi={https://doi.org/10.1175/JCLI-D-18-0661.1}
}

@article{Klotzbach2025remarkable,
  title={{The remarkable 2024 North Atlantic mid-season hurricane lull}},
  author={Klotzbach, Philip J and Bercos-Hickey, Emily and Wood, Kimberly M and Schreck III, Carl J and Bell, MM and Blake, Eric S and Bowen, Steven G and Caron, L-P and Chavas, DR and Collins, Jennifer M and others},
  journal={Geophysical Research Letters},
  volume={52},
  number={19},
  pages={e2025GL116714},
  year={2025},
  doi={https://doi.org/10.1029/2025GL116714}
}

@article{Janowiak2017ncep,
  title={{NCEP/CPC L3 half hourly 4km global (60S-60N) merged IR V1}},
  author={Janowiak, John and Joyce, Bob and Xie, Pingping},
  journal={NASA Goddard Earth Sciences Data and Information Services Center (DAAC) data set},
  pages={P4HZB9N27EKU},
  year={2017},
  doi={https://doi.org/10.5067/P4HZB9N27EKU}
}

@article{Knapp2010international,
  title={{The International Best Track Archive for Climate Stewardship (IBTrACS): Unifying tropical cyclone data}},
  author={Knapp, Kenneth R and Kruk, Michael C and Levinson, David H and Diamond, Howard J and Neumann, Charles J},
  journal={Bulletin of the American Meteorological Society},
  volume={91},
  number={3},
  pages={363--376},
  year={2010},
  doi={https://doi.org/10.1175/2009BAMS2755.1}
}

@article{Hsieh2020large,
  title={Large-scale control on the frequency of tropical cyclones and seeds: A consistent relationship across a hierarchy of global atmospheric models},
  author={Hsieh, Tsung-Lin and Vecchi, Gabriel A and Yang, Wenchang and Held, Isaac M and Garner, Stephen T},
  journal={Climate Dynamics},
  volume={55},
  number={11},
  pages={3177--3196},
  year={2020},
  doi={https://doi.org/10.1007/s00382-020-05446-5}
}

@article{Vecchi2019tropical,
  title={{Tropical cyclone sensitivities to CO2 doubling: roles of atmospheric resolution, synoptic variability and background climate changes}},
  author={Vecchi, Gabriel A and Delworth, Thomas L and Murakami, Hiroyuki and Underwood, Seth D and Wittenberg, Andrew T and Zeng, Fanrong and Zhang, Wei and Baldwin, Jane W and Bhatia, Kieran T and Cooke, William and others},
  journal={Climate Dynamics},
  volume={53},
  number={9},
  pages={5999--6033},
  year={2019},
  doi={https://doi.org/10.1007/s00382-019-04913-y}
}

@article{Berry2012african,
  title={{African Easterly Wave Dynamics in a Mesoscale Numerical Model: The Upscale Tole of Convection}},
  author={Berry, Gareth J and Thorncroft, Chris D},
  journal={Journal of the Atmospheric Sciences},
  volume={69},
  number={4},
  pages={1267--1283},
  year={2012},
  doi={https://doi.org/10.1175/JAS-D-11-099.1}
}

@article{Bercos2024characteristics,
  title={{Characteristics and trends of Atlantic tropical cyclones that do and do not develop from African easterly waves}},
  author={Bercos--Hickey, Emily and Patricola--DiRosario, Christina M},
  journal={Quarterly Journal of the Royal Meteorological Society},
  volume={150},
  number={765},
  pages={4951--4968},
  year={2024},
  doi={https://doi.org/10.1002/qj.4850}
}

@article{Chen2008north,
  title={{North Atlantic hurricanes contributed by African easterly waves north and south of the African easterly jet}},
  author={Chen, Tsing-Chang and Wang, Shih-Yu and Clark, Adam J},
  journal={Journal of Climate},
  volume={21},
  number={24},
  pages={6767--6776},
  year={2008},
  doi={https://doi.org/10.1175/2008JCLI2523.1}
}

@article{Martin2015representation,
  title={{Representation of African easterly waves in CMIP5 models}},
  author={Martin, Elinor R and Thorncroft, Chris},
  journal={Journal of Climate},
  volume={28},
  number={19},
  pages={7702--7715},
  year={2015},
  doi={https://doi.org/10.1175/JCLI-D-15-0145.1}
}

@article{Schar2020kilometer,
  title={Kilometer-scale climate models: {P}rospects and challenges},
  author={Sch{\"a}r, Christoph and Fuhrer, Oliver and Arteaga, Andrea and Ban, Nikolina and Charpilloz, Christophe and Di Girolamo, Salvatore and Hentgen, Laureline and Hoefler, Torsten and Lapillonne, Xavier and Leutwyler, David and others},
  journal={Bulletin of the American Meteorological Society},
  volume={101},
  number={5},
  pages={E567--E587},
  year={2020},
  doi={https://doi.org/10.1175/BAMS-D-18-0167.1}
}

@article{Feng2025mesoscale,
	title={{Mesoscale Convective Systems tracking Method Intercomparison (MCSMIP): Application to DYAMOND Global km-scale Simulations}},
	author={Feng, Zhe and Prein, Andreas Franz and Kukulies, Julia and Fiolleau, Thomas and Jones, William Kenneth and Maybee, Ben and Moon, Zachary and Ocasio, Kelly M N{\'u}{\~n}ez and Dong, Wenhao and Molina, Maria J and others},
	journal={Journal of Geophysical Research: Atmospheres},
    volume={130},
    pages={e2024JD042204},
	year={2025},
	doi={https://doi.org/10.1029/2024JD042204}
}

@article{Emlaw2026understanding,
  title={Understanding tropical cyclone frequency biases in a global storm resolving model},
  author={Emlaw, Geraldine Neljon and Kim, Daehyun and Blossey, Peter N and Harris, Lucas and Fueglistaler, Stephan and Moon, Jihong},
  journal={Authorea Preprints},
  year={2026},
  doi={https://doi.org/10.22541/essoar.177046458.84317593/v1}
}

@article{Baker2024realism,
  title={On the realism of tropical cyclone intensification in global storm-resolving climate models},
  author={Baker, Alexander J and Vanni{\`e}re, Beno{\^\i}t and Vidale, Pier Luigi},
  journal={Geophysical Research Letters},
  volume={51},
  number={17},
  pages={e2024GL109841},
  year={2024},
  doi={https://doi.org/10.1029/2024GL109841}
}

@article{Roberts2020impact,
  title={{Impact of model resolution on tropical cyclone simulation using the HighResMIP--PRIMAVERA multimodel ensemble}},
  author={Roberts, Malcolm John and Camp, Joanne and Seddon, Jon and Vidale, Pier Luigi and Hodges, Kevin and Vanni{\`e}re, Beno{\^\i}t and Mecking, Jenny and Haarsma, Rein and Bellucci, Alessio and Scoccimarro, Enrico and others},
  journal={Journal of Climate},
  volume={33},
  number={7},
  pages={2557--2583},
  year={2020},
  doi={https://doi.org/10.1175/JCLI-D-19-0639.1}
}

@article{Chagnon2009horizontal,
  title={Horizontal potential vorticity dipoles on the convective storm scale},
  author={Chagnon, Jeffrey M and Gray, Suzanne Louise},
  journal={Quarterly Journal of the Royal Meteorological Society},
  volume={135},
  number={643},
  pages={1392--1408},
  year={2009},
  doi={https://doi.org/10.1002/qj.468}
}

@article{Seidel2026convective,
  title={How convective mass flux responds to environmental humidity},
  author={Seidel, Seth D and Arnold, Nathan P and Wolding, Brandon},
  journal={Journal of Advances in Modeling Earth Systems},
  volume={18},
  number={2},
  pages={e2025MS005289},
  year={2026},
  doi={https://doi.org/10.1029/2025MS005289}
}

@article{Lilly1986structure,
  title={{The structure, energetics and propagation of rotating convective storms. Part I: Energy exchange with the mean flow}},
  author={Lilly, Douglas K},
  journal={Journal of Atmospheric Sciences},
  volume={43},
  number={2},
  pages={113--125},
  year={1986},
  doi={https://doi.org/10.1175/1520-0469(1986)043<0113:TSEAPO>2.0.CO;2}
}

@article{Oertel2021quantifying,
  title={Quantifying the circulation induced by convective clouds in kilometer--scale simulations},
  author={Oertel, Annika and Schemm, Sebastian},
  journal={Quarterly Journal of the Royal Meteorological Society},
  volume={147},
  number={736},
  pages={1752--1766},
  year={2021},
  doi={https://doi.org/10.1002/qj.3992}
}

@article{Lock2000new,
  title={{A new boundary layer mixing scheme. Part I: Scheme description and single-column model tests}},
  author={Lock, AP and Brown, AR and Bush, MR and Martin, GM and Smith, RNB},
  journal={Monthly Weather Review},
  volume={128},
  number={9},
  pages={3187--3199},
  year={2000},
  doi={https://doi.org/10.1175/1520-0493(2000)128<3187:ANBLMS>2.0.CO;2}
}

@article{Willett2026met,
  title={{The Met Office Unified Model Global Atmosphere 8.0 and JULES Global Land 9.0 configurations}},
  author={Willett, Martin and Brooks, Melissa and Bushell, Andrew and Earnshaw, Paul and Smith, Samantha and Tomassini, Lorenzo and Best, Martin and Boutle, Ian and Brooke, Jennifer and Edwards, John M and others},
  journal={Geoscientific Model Development},
  volume={19},
  number={4},
  pages={1473--1517},
  year={2026},
  doi={https://doi.org/10.5194/gmd-19-1473-2026}
}

@article{Boutle2016london,
  title={The {L}ondon {M}odel: forecasting fog at 333 m resolution},
  author={Boutle, Ian A and Finnenkoetter, Anke and Lock, Adrian P and Wells, H},
  journal={Quarterly Journal of the Royal Meteorological Society},
  volume={142},
  number={694},
  pages={360--371},
  year={2016},
  doi={https://doi.org/10.1002/qj.2656}
}

@article{Du2025global,
  title={Global coupled dynamics of tropical easterly waves and tropical cyclone genesis},
  author={Du, Xueqing and Chu, Jung-Eun and Jin, Fei-Fei and Cheung, Hung Ming},
  journal={npj Climate and Atmospheric Science},
  volume={8},
  number={1},
  pages={125},
  year={2025},
  doi={https://doi.org/10.1038/s41612-025-01014-y}
}

@article{Pytharoulis1999low,
  title={{The low-level structure of African Easterly Waves in 1995}},
  author={Pytharoulis, Ioannis and Thorncroft, Chris},
  journal={Monthly Weather Review},
  volume={127},
  number={10},
  pages={2266--2280},
  year={1999},
  doi={https://doi.org/10.1175/1520-0493(1999)127<2266:TLLSOA>2.0.CO;2}
}

@article{Hodges1997distribution,
  title={{Distribution and statistics of African mesoscale convective weather systems based on the ISCCP Meteosat imagery}},
  author={Hodges, Kevin and Thorncroft, CD},
  journal={Monthly Weather Review},
  volume={125},
  number={11},
  pages={2821--2837},
  year={1997},
  doi={https://doi.org/10.1175/1520-0493(1997)125<2821:DASOAM>2.0.CO;2}
}

@article{Yang2018linking,
  title={{Linking African easterly wave activity with equatorial waves and the influence of Rossby waves from the Southern Hemisphere}},
  author={Yang, Gui-Ying and Methven, John and Woolnough, Steve and Hodges, Kevin and Hoskins, Brian},
  journal={Journal of the Atmospheric Sciences},
  volume={75},
  number={6},
  pages={1783--1809},
  year={2018},
  doi={https://doi.org/10.1175/JAS-D-17-0184.1}
}

@article{Feng2023equatorial,
  title={Equatorial waves as useful precursors to tropical cyclone occurrence and intensification},
  author={Feng, Xiangbo and Yang, Gui-Ying and Hodges, Kevin and Methven, John},
  journal={Nature Communications},
  volume={14},
  number={1},
  pages={511},
  year={2023},
  doi={https://doi.org/10.1038/s41467-023-36055-5}
}

@article{Duvel2021vortices,
  title={{On vortices initiated over West Africa and their impact on North Atlantic tropical cyclones}},
  author={Duvel, Jean-Philippe},
  journal={Monthly Weather Review},
  volume={149},
  number={2},
  pages={585--601},
  year={2021},
  doi={https://doi.org/10.1175/MWR-D-20-0252.1}
}

@article{Donlon2012operational,
  title={{The operational sea surface temperature and sea ice analysis (OSTIA) system}},
  author={Donlon, Craig J and Martin, Matthew and Stark, John and Roberts-Jones, Jonah and Fiedler, Emma and Wimmer, Werenfrid},
  journal={Remote Sensing of Environment},
  volume={116},
  pages={140--158},
  year={2012},
  doi={https://doi.org/10.1016/j.rse.2010.10.017}
}

@article{Chen2014relation,
  title={{The relation between dry vortex merger and tropical cyclone genersis over the Atlantic Ocean}},
  author={Chen, Chu--Hua and Liu, Yi--Chin},
  journal={Journal of Geophysical Research: Atmospheres},
  volume={119},
  number={11},
  pages={11,641--11,661},
  year={2014},
  doi={https://doi.org/10.1002/2014JD021749}
}

@article{Cook1999generation,
	title={{Generation of the African Easterly Jet and its role in determining West African precipitation}},
	author={Cook, Kerry H},
	journal={Journal of Climate},
	volume={12},
	number={5},
	pages={1165--1184},
	year={1999},
	doi={https://doi.org/10.1175/1520-0442(1999)012<1165:GOTAEJ>2.0.CO;2}
}

@article{Hsieh2007study,
	title={{A Study of the Energetics of African Easterly Waves Using a Regional Climate Model}},
	author={Hsieh, Jen--Shan and Cook, Kerry H},
	journal={Journal of the Atmospheric Sciences},
	volume={64},
	pages={421--440},
	year={2007},
	doi={https://doi.org/10.1175/JAS3851.1}
}

@article{Semunegus2017characterization,
	title={{Characterization of convective systems and their association with African easterly waves}},
	author={Semunegus, Hilawe and Mekonnen, Ademe and Schreck III, Carl J.},
	journal={International Journal of Climatology},
	volume={37},
	pages={4486--4492},
	year={2017},
	doi={https://doi.org/10.1002/joc.5085}
}

@article{Hopsch2006west,
	title={{West African Storm Tracks and Their Relationship to Atlantic Tropical Cyclones}},
	author={Hopsch, Susanna B and Thorncrof, Chris D and Hodges, K and Aiyyer, A},
	journal={Journal of Climate},
	volume={20},
	pages={2468--2483},
	year={2006},
	doi={https://doi.org/10.1175/JCLI4139.1}
}

@article{Slingo2022ambitious,
title={Ambitious partnership needed for reliable climate prediction},
author={Slingo, Julia and Bates, Paul and Bauer, Peter and Belcher, Stephen and Palmer, Tim and Stephens, Graeme and Stevens, Bjorn and Stocker, Thomas and Teutsch, Georg},
journal={Nature Climate Change},
volume={12},
number={6},
pages={499--503},
year={2022},
doi={https://doi.org/10.1038/s41558-022-01384-8}
}

@article{Bercos2023effects,
  title={{The Effects of African Easterly Wave Suppression by Wave Track on Atlantic Tropical Cyclones}},
  author={Bercos--Hickey, Emily and Patricola, Christina M},
  journal={Geophysical Research Letters},
  volume={50},
  pages={e2023GL105491},
  year={2023},
  doi={https://doi.org/10.1029/2023GL105491}
}

@article{Asaadi2016dynamics,
  title={{On the dynamics of the formation of the Kelvin cat’s-eye in tropical cyclogenesis. Part I: Climatological investigation}},
  author={Asaadi, Ali and Brunet, Gilbert and Yau, MK},
  journal={Journal of the Atmospheric Sciences},
  volume={73},
  number={6},
  pages={2317--2338},
  year={2016},
  doi={https://doi.org/10.1175/JAS-D-15-0156.1}
}

@article{Jonville2025distinguishing,
  title={Distinguishing north and south African Easterly Waves with a spectral method: Implication for tropical cyclogenesis from mergers in the North Atlantic},
  author={Jonville, Tanguy and Cornillault, Erwan and Lavaysse, Christophe and Peyrill{\'e}, Philippe and Flamant, Cyrille},
  journal={Quarterly Journal of the Royal Meteorological Society},
  volume={151},
  number={767},
  pages={e4909},
  year={2025},
  doi={https://doi.org/10.1002/qj.4909}
}

@article{Wielicki1996clouds,
  title={{Clouds and the Earth's Radiant Energy System (CERES): An earth observing system experiment}},
  author={Wielicki, Bruce A and Barkstrom, Bruce R and Harrison, Edwin F and Lee III, Robert B and Smith, G Louis and Cooper, John E},
  journal={Bulletin of the American Meteorological Society},
  volume={77},
  number={5},
  pages={853--868},
  year={1996},
  doi={https://doi.org/10.1175/1520-0477(1996)077<0853:CATERE>2.0.CO;2}
}

@article{Martinez2025evaluating,
  title={{Evaluating environmental moisture relative to tropical deep convective growth using CPEX-CV airborne and satellite observations}},
  author={Martinez, Giselle and Rowe, Angela K and N{\'u}{\~n}ez Ocasio, Kelly M and Moon, Zachary L and Rodenkirch, Benjamin D},
  journal={Journal of Geophysical Research: Atmospheres},
  volume={130},
  number={17},
  pages={e2025JD043877},
  year={2025},
  doi={https://doi.org/10.1029/2025JD043877}
}

@article{Colon2026convective,
  title={{Convective organization in African easterly waves observed during the NAMMA and CPEX-CV field campaigns}},
  author={Col{\'o}n-Burgos, D and Bell, MM},
  journal={Journal of Geophysical Research: Atmospheres},
  volume={131},
  number={6},
  pages={e2025JD045516},
  year={2026},
  doi={https://doi.org/10.1029/2025JD045516}
}

@article{Flamant2024cyclogenesis,
  title={{Cyclogenesis in the Tropical Atlantic: First Scientific Highlights from the Clouds--Atmospheric Dynamics--Dust Interactions in West Africa (CADDIWA) Field Campaign}},
  author={Flamant, Cyrille and Chaboureau, Jean-Pierre and Delano{\"e}, Julien and Gaetani, Marco and Jamet, C{\'e}dric and Lavaysse, Christophe and Bock, Olivier and Borne, Maurus and Cazenave, Quitterie and Coutris, Pierre and others},
  journal={Bulletin of the American Meteorological Society},
  volume={105},
  number={2},
  pages={E387--E417},
  year={2024},
  doi={https://doi.org/10.1175/BAMS-D-23-0230.1}
}

@article{Mekonnen2006analysis,
  title={Analysis of convection and its association with {A}frican easterly waves},
  author={Mekonnen, Ademe and Thorncroft, Chris D and Aiyyer, Anantha R},
  journal={Journal of Climate},
  volume={19},
  number={20},
  pages={5405--5421},
  year={2006},
  doi={https://doi.org/10.1175/JCLI3920.1}
}

@article{Hopsch2010analysis,
  title={Analysis of {A}frican easterly wave structures and their role in influencing tropical cyclogenesis},
  author={Hopsch, Susanna B and Thorncroft, Chris D and Tyle, Kevin R},
  journal={Monthly Weather Review},
  volume={138},
  number={4},
  pages={1399--1419},
  year={2010},
  doi={https://doi.org/10.1175/2009MWR2760.1}
}

@article{Lavender2026comorph,
  title={{Modifying CoMorph--A for kilometre--scale resolutions}},
  author={S.L. Lavender and A.J. Stirling and S. Smith and M. Whitall and A. P. Lock},
  journal={Weather and Forecasting},
  year={2026},
  doi={https://doi.org/10.1175/WAF-D-25-0242.1}
}

@article{Luschen2026stratiform,
  title={Stratiform and anvil cloud-radiative forcing in tropical cyclogenesis},
  author={Luschen, Emily and Ruppert, James and Rios--Berrios, Rosimar and Wu, Shun--Nan and Zhang, Yunji},
  journal={Geophysical Research Letters},
  volume={53},
  number={1},
  pages={e2025GL119765},
  year={2026},
  doi={https://doi.org/10.1029/2025GL119765}
}

@article{Klotzbach2022hyperactive,
  title={{A hyperactive end to the Atlantic hurricane season October--November 2020}},
  author={Klotzbach, Philip J and Wood, Kimberly M and Bell, Michael M and Blake, Eric S and Bowen, Steven G and Caron, Louis-Philippe and Collins, Jennifer M and Gibney, Ethan J and Schreck III, Carl J and Truchelut, Ryan E},
  journal={Bulletin of the American Meteorological Society},
  volume={103},
  number={1},
  pages={E110--E128},
  year={2022},
  doi={https://doi.org/10.1175/BAMS-D-20-0312.1}
}

@article{Prein2021sensitivity,
  title={Sensitivity of organized convective storms to model grid spacing in current and future climates},
  author={Prein, Andreas F and Rasmussen, Roy M and Wang, Di{\'e} and Giangrande, Scott E},
  journal={Philosophical Transactions of the Royal Society A: Mathematical, Physical and Engineering Sciences},
  volume={379},
  number={2195},
  year={2021},
  doi={ https://doi.org/10.1098/rsta.2019.0546}
}

@article{Kukulies2024simulating,
  title={Simulating precipitation efficiency across the deep convective gray zone},
  author={Kukulies, Julia and Prein, Andreas F and Morrison, Hugh},
  journal={Journal of Geophysical Research: Atmospheres},
  volume={129},
  number={24},
  pages={e2024JD041924},
  year={2024},
  doi={https://doi.org/10.1029/2024JD041924}
}

@article{Lebo2015effects,
  title={Effects of horizontal and vertical grid spacing on mixing in simulated squall lines and implications for convective strength and structure},
  author={Lebo, ZJ and Morrison, H},
  journal={Monthly Weather Review},
  volume={143},
  number={11},
  pages={4355--4375},
  year={2015},
  doi={https://doi.org/10.1175/MWR-D-15-0154.1}
}

@article{Raymond2007evolution,
  title={Evolution of convection during tropical cyclogenesis},
  author={Raymond, David J and Sessions, Sharon L},
  journal={Geophysical Research Letters},
  volume={34},
  number={6},
  year={2007},
  doi={https://doi.org/10.1029/2006GL028607}
}

@article{Raymond2007theory,
  title={A theory for the spinup of tropical depressions},
  author={Raymond, David J and Sessions, Sharon L and Fuchs, {\v{Z}}eljka},
  journal={Quarterly Journal of the Royal Meteorological Society},
  volume={133},
  number={628},
  pages={1743--1754},
  year={2007},
  publisher={https://doi.org/10.1002/qj.125}
}

@article{Bister1997genesis,
  title={{The genesis of Hurricane Guillermo: TEXMEX analyses and a modeling study}},
  author={Bister, Marja and Emanuel, Kerry A},
  journal={Monthly Weather Review},
  volume={125},
  number={10},
  pages={2662--2682},
  year={1997},
  doi={https://doi.org/10.1175/1520-0493(1997)125<2662:TGOHGT>2.0.CO;2}
}

@article{Segura2025nextgems,
  title={{nextGEMS: entering the era of kilometer-scale Earth system modeling}},
  author={Segura, Hans and Pedruzo-Bagazgoitia, Xabier and Weiss, Philipp and M{\"u}ller, Sebastian K and Rackow, Thomas and Lee, Junhong and Dolores-Tesillos, Edgar and Benedict, Imme and Aengenheyster, Matthias and Aguridan, Razvan and others},
  journal={Geoscientific Model Development},
  volume={18},
  number={20},
  pages={7735--7761},
  year={2025},
  doi={https://doi.org/10.5194/gmd-18-7735-2025}
}

@article{Wing2016role,
  title={Role of radiative--convective feedbacks in spontaneous tropical cyclogenesis in idealized numerical simulations},
  author={Wing, Allison A and Camargo, Suzana J and Sobel, Adam H},
  journal={Journal of the Atmospheric Sciences},
  volume={73},
  number={7},
  pages={2633--2642},
  year={2016},
  doi={https://doi.org/10.1175/JAS-D-15-0380.1}
}

@article{Nolan2007tropical,
  title={Tropical cyclogenesis sensitivity to environmental parameters in radiative--convective equilibrium},
  author={Nolan, David S and Rappin, Eric D and Emanuel, Kerry A},
  journal={Quarterly Journal of the Royal Meteorological Society},
  volume={133},
  number={629},
  pages={2085--2107},
  year={2007},
  doi={https://doi.org/10.1002/qj.170}
}

@article{Camargo2007use,
  title={Use of a genesis potential index to diagnose {ENSO} effects on tropical cyclone genesis},
  author={Camargo, Suzana J and Emanuel, Kerry A and Sobel, Adam H},
  journal={Journal of Climate},
  volume={20},
  number={19},
  pages={4819--4834},
  year={2007},
  doi={https://doi.org/10.1175/JCLI4282.1}
}

@article{Tippett2011poisson,
  title={A {P}oisson regression index for tropical cyclone genesis and the role of large-scale vorticity in genesis},
  author={Tippett, Michael K and Camargo, Suzana J and Sobel, Adam H},
  journal={Journal of Climate},
  volume={24},
  number={9},
  pages={2335--2357},
  year={2011},
  doi={https://doi.org/10.1175/2010JCLI3811.1}
}

@article{Illingworth2015earthcare,
  title={{The EarthCARE satellite: The next step forward in global measurements of clouds, aerosols, precipitation, and radiation}},
  author={Illingworth, Anthony J and Barker, HW and Beljaars, A and Ceccaldi, Marie and Chepfer, H and Clerbaux, Nicolas and Cole, J and Delano{\"e}, Julien and Domenech, C and Donovan, David P and others},
  journal={Bulletin of the American Meteorological Society},
  volume={96},
  number={8},
  pages={1311--1332},
  year={2015},
  doi={https://doi.org/10.1175/BAMS-D-12-00227.1}
}

@article{Roh2025vertical,
  title={{Vertical motions in clouds from EarthCare satellite and a global storm-resolving modeling}},
  author={Roh, Woosub and Satoh, Masaki and Matsugishi, Shuhei and Aoki, Shunsuke and Kubota, Takuji and Okamoto, Hajime},
  journal={Scientific Reports},
  year={2025},
  doi={https://doi.org/10.1038/s41598-025-32256-8}
}

@article{Dolan2023time,
  title={Time resolved reflectivity measurements of convective clouds},
  author={Dolan, Brenda and Kollias, Pavlos and van den Heever, Susan C and Rasmussen, Kristen L and Oue, Mariko and Luke, Edward and Lamer, Katia and Treserras, Bernat P and Haddad, Ziad and Stephens, Graeme and others},
  journal={Geophysical Research Letters},
  volume={50},
  number={22},
  pages={e2023GL105723},
  year={2023},
  doi={https://doi.org/10.1029/2023GL105723}
}

@article{Wu2024contribution,
  title={{The contribution of Mesoscale Convective Systems to the coastal rainfall maximum over West Africa}},
  author={Wu, Shun-Nan and Sakaeda, Naoko and Martin, Elinor and Rios-Berrios, Rosimar and Russell, James},
  journal={Monthly Weather Review},
  volume={152},
  number={8},
  pages={1787--1802},
  year={2024},
  doi={https://doi.org/10.1175/MWR-D-23-0148.1}
}

@article{Janiga2014convection,
  title={{Convection over tropical Africa and the east Atlantic during the West African monsoon: Regional and diurnal variability}},
  author={Janiga, Matthew A and Thorncroft, Chris D},
  journal={Journal of Climate},
  volume={27},
  number={11},
  pages={4159--4188},
  year={2014},
  doi={https://doi.org/10.1175/JCLI-D-13-00449.1}
}

@article{Serra2010tracking,
  title={{Tracking and mean structure of easterly waves over the Intra-Americas Sea}},
  author={Serra, Yolande L and Kiladis, George N and Hodges, Kevin I},
  journal={Journal of Climate},
  volume={23},
  number={18},
  pages={4823--4840},
  year={2010},
  doi={https://doi.org/10.1175/2010JCLI3223.1}
}

@article{Dunkerton2009tropical,
  title={Tropical cyclogenesis in a tropical wave critical layer: {E}asterly waves},
  author={Dunkerton, Timothy J and Montgomery, MT and Wang, Z},
  journal={Atmospheric Chemistry and Physics},
  volume={9},
  number={15},
  pages={5587--5646},
  year={2009},
  doi={https://doi.org/10.5194/acp-9-5587-2009}
}

@article{Takasuka2026precipitation,
  title={Precipitation characteristics and thermodynamic-convection coupling in global kilometer-scale simulations},
  author={Takasuka, Daisuke and Becker, Tobias and Bao, Jiawei},
  journal={Journal of Advances in Modeling Earth Systems},
  volume={18},
  number={3},
  pages={e2025MS005343},
  year={2026},
  doi={https://doi.org/10.1029/2025MS005343}
}

@article{Rios2024modulation,
  title={{Modulation of Tropical Cyclogenesis by Convectively Coupled Kelvin Waves}},
  author={Rios--Berrios, Rosimar and Tang, Brian H and Davis, Christopher A and Martinez, Jonathan},
  journal={Monthly Weather Review},
  volume={152},
  number={10},
  pages={2309--2322},
  year={2024},
  doi={https://doi.org/10.1175/MWR-D-24-0052.1 }
}

@article{Lawton2022influence,
  title={{The influence of Convectively Coupled Kelvin Waves on African Easterly Waves in a wave-following framework}},
  author={Lawton, Quinton A and Majumdar, Sharanya J and Dotterer, Krista and Thorncroft, Christopher and Schreck III, Carl J},
  journal={Monthly Weather Review},
  volume={150},
  number={8},
  pages={2055--2072},
  year={2022},
  doi={https://doi.org/10.1175/MWR-D-21-0321.1}
}

@article{Roberts2015tropical,
  title={{Tropical cyclones in the UPSCALE ensemble of high-resolution global climate models}},
  author={Roberts, Malcolm J and Vidale, Pier Luigi and Mizielinski, Matthew S and Demory, Marie-Estelle and Schiemann, Reinhard and Strachan, Jane and Hodges, Kevin and Bell, Ray and Camp, Joanne},
  journal={Journal of Climate},
  volume={28},
  number={2},
  pages={574--596},
  year={2015},
  doi={https://doi.org/10.1175/JCLI-D-14-00131.1}
}

@article{Wang2014characteristics,
  title={Characteristics of tropical easterly wave pouches during tropical cyclone formation},
  author={Wang, Zhuo and Hankes, Isaac},
  journal={Monthly Weather Review},
  volume={142},
  number={2},
  pages={626--633},
  year={2014},
  doi={https://doi.org/10.1175/MWR-D-13-00267.1}
}

@article{Leppert2013relation,
  title={Relation between tropical easterly waves, convection, and tropical cyclogenesis: {A} {L}agrangian perspective},
  author={Leppert, Kenneth D and Cecil, Daniel J and Petersen, Walter A},
  journal={Monthly Weather Review},
  volume={141},
  number={8},
  pages={2649--2668},
  year={2013},
  doi={https://doi.org/10.1175/MWR-D-12-00217.1}
}

@article{Maybee2026zenodo,
  title={{BMaybee/East\_Atlantic\_vortices}},
  author={Maybee, Ben},
  journal={Zenodo},
  year={2026},
  doi={https://doi.org/10.5281/zenodo.19860489}
}

@article{Nunez2026review,
	title={{A Review of African Easterly Waves: Formation, Growth, Structure, and Evolution}},
	author={Nu\~{n}ez Oc\'{a}sio, Kelly M and Lawton, Quinton A and Brammer, Alan and Bercos--Hickey, Emily and Bourdin, Stella and Feng, Xiangbo and Schwendike, Juliane},
	journal={Journal of the Atmospheric Sciences},
	volume={83},
	number={6},
	pages={901--926},
	year={2026},
	doi={https://doi.org/10.1175/JAS-D-25-0175.1}
}

@article{Ullrich2021tempestextremes,
  title={{TempestExtremes v2. 1: A community framework for feature detection, tracking, and analysis in large datasets}},
  author={Ullrich, Paul A and Zarzycki, Colin M and McClenny, Elizabeth E and Pinheiro, Marielle C and Stansfield, Alyssa M and Reed, Kevin A},
  journal={Geoscientific Model Development},
  volume={14},
  number={8},
  pages={5023--5048},
  year={2021},
  doi={https://doi.org/10.5194/gmd-14-5023-2021}
}

@article{Bell2000climate,
  title={Climate assessment for 1999},
  author={Bell, Gerald D and Halpert, Michael S and Schnell, Russell C and Higgins, R Wayne and Lawrimore, Jay and Kousky, Vernon E and Tinker, Richard and Thiaw, Wasila and Chelliah, Muthuvel and Artusa, Anthony},
  journal={Bulletin of the American Meteorological Society},
  volume={81},
  number={6},
  pages={S1--S50},
  year={2000},
  doi={https://doi.org/10.1175/1520-0477(2000)81[s1:CAF]2.0.CO;2}
}

@article{Garcia2013impact,
  title={The impact of convective cold pool outflows on model biases in the Sahara},
  author={Garcia--Carreras, L and Marsham, JH and Parker, DJ and Bain, CL and Milton, S and Saci, A and Salah-Ferroudj, M and Ouchene, B and Washington, R},
  journal={Geophysical Research Letters},
  volume={40},
  number={8},
  pages={1647--1652},
  year={2013},
  doi={https://doi.org/10.1002/grl.50239}
}

@article{Gehne2022diagnostics,
  title={Diagnostics of tropical variability for numerical weather forecasts},
  author={Gehne, Maria and Wolding, Brandon and Dias, Juliana and Kiladis, George N},
  journal={Weather and Forecasting},
  volume={37},
  number={9},
  pages={1661--1680},
  year={2022},
  doi={https://doi.org/10.1175/WAF-D-21-0204.1}
}

@article{Grist2001study,
  title={{A study of the dynamic factors influencing the rainfall variability in the West African Sahel}},
  author={Grist, Jeremy P and Nicholson, Sharon E},
  journal={Journal of Climate},
  volume={14},
  number={7},
  pages={1337--1359},
  year={2001},
  doi={https://doi.org/10.1175/1520-0442(2001)014<1337:ASOTDF>2.0.CO;2}
}

\end{document}